\documentclass[12pt,apj]{emulateapj}

\usepackage{subfigure}
\usepackage[usenames]{color}

\begin{document}

\shorttitle{Impact of partial ionization on the solar atmosphere}
\shortauthors{Mart\'inez-Sykora et al.}
\title{Two-dimensional Radiative Magnetohydrodynamic Simulations of Partial Ionization in the
  Chromosphere. II. Dynamics and Energetics of the Low Solar Atmosphere}

\author{Juan Mart\'inez-Sykora\altaffilmark{1,2}}
\email{juanms@lmsal.com}
\author{Bart De Pontieu\altaffilmark{2,3}}
\author{Mats Carlsson\altaffilmark{3}}
\author{Viggo H. Hansteen\altaffilmark{3,2}}
\author{Daniel N\'obrega-Siverio\altaffilmark{4,5}}
\author{Boris V. Gudiksen\altaffilmark{3}}

\affil{\altaffilmark{1} Bay Area Environmental Research Institute, Petaluma, CA 94952, USA}
\affil{\altaffilmark{2} Lockheed Martin Solar and Astrophysics Laboratory, Palo Alto, CA 94304, USA}
\affil{\altaffilmark{3} Institute of Theoretical Astrophysics, University of Oslo, P.O. Box 1029 Blindern, N-0315 Oslo, Norway}
\affil{\altaffilmark{4} Instituto de Astrof\'isica de Canarias, 38200 La Laguna (Tenerife), Spain} 
\affil{\altaffilmark{5} Department of Astrophysics, Universidad de La Laguna, E-38200 La Laguna (Tenerife), Spain}

\newcommand{\eg}{{\it e.g.,}} 
\newcommand{\myemail}{juanms@astro.uio.no}
\newcommand{\komment}[1]{\texttt{#1}}

\begin{abstract}
We investigate the effects of interactions between ions and neutrals on the 
chromosphere and overlying corona using 2.5D radiative MHD 
simulations with the {\it Bifrost} code. We have extended the code capabilities implementing ion-neutral interaction 
effects using the Generalized Ohm's Law, {\it i.e.}, we include the Hall term 
and the ambipolar diffusion (Pedersen dissipation) in the induction equation. 
Our models span from the upper convection zone to the corona, with the photosphere, chromosphere 
and transition region partially ionized. Our simulations
reveal that the interactions between ionized particles 
and neutral particles have important consequences for the
magneto-thermodynamics of these modeled layers: 1) ambipolar diffusion 
increases the temperature in the chromosphere; 2) sporadically the horizontal magnetic field 
in the photosphere is diffused into the chromosphere due to the large ambipolar 
diffusion; 3) ambipolar diffusion concentrates electrical currents leading to more violent 
jets and reconnection processes, resulting in 3a) the formation of
longer and faster spicules, 3b) heating of plasma during the spicule evolution,
and 3c) decoupling of the plasma and magnetic field in spicules. 
Our results indicate that ambipolar diffusion is a critical ingredient
for understanding the magneto-thermo-dynamic properties in the
chromosphere and transition region. 
The numerical simulations have been made publicly
   available, similar to previous Bifrost
   simulations. This will allow the community to study realistic numerical
   simulations with a wider range
   of magnetic field configurations and physics modules than previously
   possible.

\end{abstract}


\section{Introduction}

The chromosphere (the interface between the photosphere and 
the million degree corona) is of great interest because it processes
enough non-thermal energy to heat the entire corona, and all
non-thermal energy that powers the corona and solar wind passes first
through the chromosphere. The chromosphere thus plays a key role in the energy and mass balance of 
the outer solar atmosphere \citep{Athay:1982fk,Dere:1989uq,De-Pontieu:2011lr}, 
and is at the root of the solar wind 
\citep{Withbroe:1977kx,De-Pontieu:2007bd,Tomczyk:2007vn,McIntosh:2011fk}. 
However, it is a very complex region that is difficult to directly
diagnose. This is because many complex physical processes play a role in the 
chromosphere: 1) the plasma the plasma is not in Local Thermodynamic Equilibrium (LTE), 2) 
the radiation is optically thick, 3) the radiation suffers scattering, 4) the ionization is not 
in equilibrium, 5) the gas is partially ionized, 6) in the upper chromosphere, transition 
region (TR) and corona, thermal conduction plays an important role in
transporting energy along the magnetic field. 
In addition, the chromosphere is also distinguished by several
transitions: from gas pressure dominated to magnetic field dominated,
from collisional to collisionless, and from partial 
to full ionization. Note that many of the physical processes
(specially 1-4) imply that the interpretation of imaging and spectral
observations is often difficult and requires modeling.

The chromosphere is highly dynamic: it is permeated by upward
traveling shocks \citep{Hansteen+DePontieu2006,De-Pontieu:2007kl} and its interface with the TR and corona is dominated by short lived jets or spicules. Many phenomena in the chromosphere remain 
poorly understood. For example, it is not clear
how the chromosphere is heated in both non-magnetic regions
\citep{Carlsson:2007uq} and magnetic regions \citep{Carlsson:2015fk}. 
It also remains unclear how small-scale flux emerges into the chromosphere:  the
chromosphere appears to be filled with magnetic field, even in quiet
Sun, despite the fact that the photosphere is sub-adiabatic which
should, in principle, thwart the 
expansion of emerging magnetic flux into the corona \citep{Acheson:1979lr,archontis2004}.
The formation of spicules has similarly remained mysterious with a multitude of models proposed \citep{Sterling:2010vn,Tsiropoula:2012yq}, but most failing to reproduce the properties of spicules as they are now measured with high-resolution instruments \citep{de-Pontieu:2007wx,Pereira:2012dz}; for example these models fail to reproduce the very high speeds ($50$~km~s$^{-1}$ or more), the temperature evolution of observed chromospheric spicules or the short lived Rapid Blue Events seen in the spectral profiles of \ion{Ca}{2}, or H$\alpha$ \citep{Martinez-Sykora:2013ys}, or the heating to transition region temperatures. Recently,
\citet{Martinez-Sykora:2017sci} used the same model as the current 
paper to propose a new spicule formation mechanism that can reproduce
the observed properties of spicules. This model has the potential to
resolve some of the remaining unresolved issues with respect to the
impact of spicules on the transition region and corona
\citep{De-Pontieu:2011lr,Madjarska:2011fk,Judge:2012lh}.
Other phenomena are better understood, but current models cannot fully capture 
some of their properties, {\it e.g.}, the length and lifetime of dynamic fibrils 
\citep{Suematsu:1995lr,Hansteen+DePontieu2006,Heggland:2007jt,
Martinez-Sykora:2009kl,Iijima:2015fk}. 
The persistence of these unresolved issues in part stems from the difficulty of combining all
these ingredients into a single numerical approach  that models the
full solar atmosphere and takes into account radiative transfer with scattering, thermal conduction, partial ionization effects, etc. 

Significant efforts have focused on combining 3D MHD equations with radiative transfer 
and thermal conduction along the magnetic field 
\citep{Stein:2006qy,abbett2007,Gudiksen:2011qy,Bingert:2011fk,Rempel:2017zl}. 
Furthermore, recently \citet{Leenaarts:2007sf} and \citet{Golding:2014fk} expanded the {\it Bifrost} code to 
include non-equilibrium ionization. \citet{Cheung:2012uq} implemented the Hall term 
in photospheric simmulations of the 3D radiative MHD {\it MURaM} code.
In this paper we use a version of the {\it Bifrost} code that includes the Hall term 
and Pedersen dissipation in an atmosphere that spans from the convection zone to 
the lower corona. Our first results \citep{Martinez-Sykora:2012uq} focused on the 
validity of the generalized Ohm's law and on the spatio-temporal properties of the Hall term and the ambipolar
diffusion in a 2D radiative MHD solar atmosphere. Here we will focus
on the impact of these terms on the magneto-thermo-dynamic properties
of the solar atmosphere.

In a magnetized partially ionized gas, ions are coupled to the magnetic field, whereas neutrals are not directly affected by the magnetic field and, in principle, can move ``freely". This can lead to a velocity drift between ions and neutrals, consequently the bulk motion of the combined fluid can, under certain circumstances, be different from the motion of the magnetic flux 
\citep{Martinez-Sykora:2016qf}. Sufficient collisions between ions and neutrals can couple the neutrals to the magnetic field, while at the same time to some extent decouple the ions from the magnetic field. Under certain conditions, i.e., when collisions are sufficient to ensure equal temperatures of the ions, neutrals and electrons , and in addition that timescales are greater than the ion-neutral collision timescales, one can still solve the single fluid MHD equations and expand Ohm’s law in order to include the ion-neutral interaction effects, by adding the so-called Hall term and ambipolar diffusion \citep[see][amongst others]{cowling1957,Braginskii:1965ul,Parker:2007lr}.
As a result of ambipolar diffusion, the magnetic field can diffuse and magnetic energy will be dissipated and lead to heating.  
These are the effects implemented in the 3D radiative MHD {\it Bifrost} code (Section~\ref{sec:equations}). 

During the past two decades significant progress has been made in understanding 
the potential impact of partial ionization in the lower solar atmosphere, 
mostly using more idealized models. Partial ionization in the chromosphere is for 
instance known to lead to dissipation of Alfv\'enic waves through ion-neutral 
collisions \citep[][among others]{de-Pontieu:1998lr,de-Pontieu:1999kx, 
De-Pontieu:2001fj,Leake:2005rt,Forteza:2007zp,Soler:2009hw,Soler:2012hb,Soler:2013la,Soler:2015gd}. 
However, the picture is far from complete since it remains unknown if this dissipation 
plays a significant role in heating the chromosphere: previous work
was based on highly idealized models (usually assuming a static or
structure-less chromosphere) and most often used fixed values for
ion-neutral collision frequencies, ignoring the intricate dynamic
balance between heating, ionization and cooling that continuously
takes place in the chromosphere.
Previous work has also found that electrical currents perpendicular to the magnetic field can be dissipated by Pedersen 
dissipation and lead to heating \citep[{\it e.g.},][]{Arber:2009ve,Khomenko:2012bh,Goodman:2012vn}. Pedersen 
dissipation also allows emerging magnetic field to diffuse more rapidly into the atmosphere 
\citep{Leake2006,Arber:2007yf,Leake:2013dq}. Compared to single fluid
simulations, 2.5D simulations of prominences 
including partial ionization show an increase of small scale velocities as a result of
the non-linearity of the Rayleigh-Taylor instability \citep{Diaz:2014dq,Khomenko:2014yg}. Further details of such processes 
can be found in the review by \citet{Leake:2014fk}, which details the impact of ion-neutral
collisions, and their properties in the solar chromosphere and the Earth's ionosphere. 
\cite{Martinez-Sykora:2015lq} summarize and discuss the role of ion-neutral interaction 
effects in the solar atmosphere.

This paper is structured as follows: in Section~\ref{sec:equations} we briefly describe
the {\it Bifrost} code \citep{Gudiksen:2011qy}, as well as the numerical methods used to model the ion-neutral 
interaction effects. 
We study the impact of ion-neutral effects by studying two
simulations, one without and one with ion-neutral interaction effects. The initial conditions and the setup of the two simulations are detailed in 
Section~\ref{sec:models}.  In
  Section~\ref{sec:dataaccess} we provide details on the publicly
  available snapshots from the simulations. We analyze these models in detail in
Section~\ref{sec:res}, where we focus on general aspects of the thermodynamic properties 
(Section~\ref{sec:termo}), the magnetic field distribution (Section~\ref{sec:magnetic}), the heating  (Section~\ref{sec:heating})
and the transport of magnetic flux (Section~\ref{sec:flux}).
We continue
with a description of three representative processes that we consider of 
great interest (Section~\ref{sec:processes}): expanding cold chromospheric bubbles (Section~\ref{sec:shocks}),  
chromospheric jets (Section~\ref{sec:jets}), and chromospheric reconnection in regions 
with highly inclined field (Section~\ref{sec:chromjets}).
We finish the paper with a discussion and conclusions in Section~\ref{sec:conclusions}. 

\section{Equations and Numerical Method}~\label{sec:equations}

The photosphere, chromosphere, and transition region are partially ionized
and the interaction between ionized and neutral particles has important 
consequences. We investigate this physical process by modeling the solar 
atmosphere with the {\it Bifrost} code. This code solves the full MHD equations 
with radiative transfer and thermal conduction along the magnetic field. 
The numerical methods implemented in the code have been described
in detail by \citet{Gudiksen:2011qy}. In addition, the modules for the optically 
thick radiation and the numerical recipes for the radiative transfer in the 
chromosphere and TR are described by \citet{Hayek:2010ac} 
and \citet{Carlsson:2012uq}, respectively. In order to implement the ion-neutral interaction  
effects we take into account that the code explicitly solves the MHD 
equations on a Cartesian staggered mesh. In order to suppress numerical 
noise, high-order artificial diffusion is added both in the form of 
viscosity and magnetic diffusivity 
\citep[see][]{mhdcodeboris,Gudiksen:2011qy,Martinez-Sykora:2009rw}. 

We implemented partial ionization effects in the {\it Bifrost} code by adding two 
new terms in the induction equation, {\it i.e.}, the Hall term and the ambipolar diffusion:

\begin{eqnarray}
\frac{\partial {\bf B}}{\partial t} = && \nabla \times [{\bf u \times B} -  \eta {\bf J}
 - \frac{\eta_{hall}}{ |B|} {\bf J \times B} +  \nonumber \\ 
&& \frac{\eta_{amb}}{ B^2} ({\bf J \times B}) \times {\bf B} ]\label{eq:faradtot2}
\end{eqnarray}

\noindent
where ${\bf B}$, ${\bf J}$, ${\bf u}$, and $\eta$ are magnetic field, current density, 
velocity, and the ohmic diffusion, respectively \citep[see][ for the derivation of this 
equation]{cowling1957,Braginskii:1965ul,Parker:2007lr,Pandey:2008qy,
Martinez-Sykora:2012uq,Leake:2014fk}. The {\it Bifrost} code 
does not calculate the Spitzer ohmic dissipation since the artificial diffusion plays this role and 
is much larger than the Spitzer ohmic diffusion. The $\eta_{hall}$ term and the $\eta_{amb}$ diffusive term are: 

\begin{eqnarray}
\eta_{hall} & = & \frac{ |B|}{q_e n_e}\\
\eta_{amb} & = & \frac{(|B| \rho_n/\rho)^2}{\rho_i \nu_{in}} = \frac{(|B|
  \rho_n/\rho)^2}{\rho_n \nu_{ni}}
\end{eqnarray}

\noindent
where $\rho_i$, $\rho_n$, $\rho$, $\nu_{in}$, $\nu_{ni}$, $n_{e}$, and $q_e$
are ion mass density, neutral mass density, total mass density, ion-neutral collision frequency, 
neutral-ion collision frequency, electron number density and the absolute value of the electron charge, 
respectively. Note that $\eta_{hall}$ is not a diffusive or dissipative term, in contrast 
to $\eta$ and $\eta_{amb}$. The ambipolar diffusion should not be confused with 
the so-called ambipolar drift used in plasma physics. Ambipolar diffusion is also 
referred to as Pedersen dissipation. Since we are using the mathematical expression 
of the ambipolar diffusion \citep[see][]{Parker:2007lr} instead of the Pedersen 
dissipation \citep[compare with Eq. 46 in][]{Leake:2014fk} in the code, we refer to this physical process as ambipolar diffusion. 

We implement four different approximations 
to compute the ion-neutral collision frequency. Three of them have been 
already discussed in \citet{Martinez-Sykora:2012uq}: one
approximation follows \citet{Osterbrock:1961fk} and \citet{de-Pontieu:1998lr}; 
the second one follows \citet{von-Steiger:1989uq}; and the third one 
follows \citet{Fontenla:1993fj}. The fourth approximation to compute 
the ion-neutral collision frequency is using recent studies that 
improve the estimation of the collisional cross sections 
under chromospheric conditions \citep{Vranjes:2013ve}. 
These various approximations for the calculation of 
the collision frequency between ions and neutrals lead to large
differences in the values of ambipolar diffusion as shown by 
\citet{Martinez-Sykora:2012uq}. 
These differences also lead to different 
thermal properties of the solar atmosphere \citep{Martinez-Sykora:2015lq}. Therefore, 
it is critical to properly calculate the ambipolar diffusion. As a result, 
we present here only results based on the most recent and state-of-the-art 
approximation of the ion-neutral cross sections described by 
\citet{Vranjes:2013ve}. For this, we take into account the temperature dependence of the 
cross sections for p-H, p-He and He$^+$-He collisions. The cross sections are calculated by combining quantum and classical theory. In addition, we consider the 
sixteen most important elements. The collisional cross sections of these elements are not well-known. 
As an approximation and following \citet{Vranjes:2008uq}, the cross section between any other 
element with neutrals is chosen to be the value of the 
cross section for protons multiplied by $m_m/m_p$, where $m_m$ is the atomic mass of the 
considered element and $m_p$ is the proton mass.

To solve the Hall term and the ambipolar diffusion when evaluating the
partial differential equations, we apply a method similar to that used in the {\it Bifrost} code,
{\it i.e.}, a sixth order accurate operator for determining the partial spatial derivatives. 

Since we can reformulate Equation~\ref{eq:faradtot2} as follows:
 
\begin{eqnarray}
\frac{\partial {\bf B}}{\partial t} =  \nabla \times \left[{\bf u \times B} - \eta {\bf J}
 - {\bf u_H \times B} +{\bf u_A \times B}\right]\label{eq:faradtot3}
\end{eqnarray}

\noindent where the ``Hall velocity" is ${\bf u_H}=(\eta_{hall}{\bf J})/|B| $ 
and the ``ambipolar velocity" is ${\bf u_A}=(\eta_{amb}{\bf J \times B})/B^2 $,
the Hall term and ambipolar diffusivity impose two new constraints
on the CFL  
condition \citep{Courant:1928uq} which restrict the time interval between numerical steps 
($\Delta t_{H}= \Delta x/{\bf u_{H}}$ and $\Delta t_{A}= \Delta x/{\bf u_{A}}$). 
Both velocities are a function of the current ($\nabla\times\mathbf{B}$), {\it i.e.}, both 
CFL conditions are quadratic functions in $\Delta x$, and the timestep will therefore
decrease quadratically with increasing spatial resolution \citep{Cheung:2012uq}. 
In addition to these CFL conditions the whistler phase speed ($u_w = k u_a^2/\Omega_i$, where
$u_a$ is the Alfv\'en velocity, $\Omega_i$ is the ion cyclotron velocity, and $k$ the 
wavenumber) is also required for completeness of the CFL 
condition for the Hall term \citep{Huba:2003fk}.

These new CFL restrictions only apply to the induction equation, which contains the 
Hall and ambipolar terms. Therefore, we solve the induction equation on a 
separate time scale, $N_{Ist} = \Delta t_{MHD}/\Delta t_{GOL}$
times for each time that the MHD equations are solved, where $\Delta t_{MHD}$ 
is the smallest timestep interval due to the classical CFL condition on the MHD equations
and $\Delta t_{GOL}=\min(\Delta t_{H},\Delta t_{A})$ is the timestep for the 
induction equation. \citep[{\it i.e.}, following a similar approach as][]{Leake:2006kx}. 
The result from solving the induction equation $N_{Ist}$ times feeds the 
MHD equations with an updated magnetic field and Joule heating 
coming from the artificial and ambipolar dissipation. 
If during the $N_{Ist}$ iterations 
the ratio between the internal energy and the accumulative magnetic energy release 
is larger than a certain threshold (${e/Q_{Joule}}>\Delta t_{MHD}$) we allow the iterations 
to be interrupted and the MHD equations to be solved again. This constraint is based on the magnitude 
of the magnetic energy release from the ambipolar diffusion. It can be calculated 
accurately since the induction equation advances in time explicitly and the 
accumulated dissipated magnetic energy is calculated for each small timestep. 

Numerical errors coming from the two new terms in the induction equation 
are diffused away using hyper-diffusive operators. Taking into account  
Equation~(\ref{eq:faradtot3}), we add a hyper-diffusive term similar to that  
described in \citet{Gudiksen:2011qy} \citep[and previously][]{mhdcodeboris} to the advection term 
($\nabla\times\mathbf{u}\times\mathbf{B}$) in the induction equation. 

\begin{eqnarray}
\frac{\partial {\bf B}}{\partial t} && =  ... +  \frac{\partial}{\partial x} \{[\max(\nu_{h1}|{\bf u_{H}}|, \nu_{a1}|{\bf u_{A}}|) +  \\
	&& \max(\nu_{h2} \Delta x \nabla_{x}^{1}u_{Hx}, \nu_{a2}\Delta x \nabla_{x}^{1}u_{Ax})] \frac{\partial {\bf B}}{\partial x} 
	Q\left(\frac{\partial {\bf B}}{\partial x}\right) \} \nonumber	
\end{eqnarray}

\noindent where $\nabla_{x}^{1}$ is the first order gradient in the $x$ direction, $\nu_{a1}$, 
$\nu_{a2}$, $\nu_{h1}$, and $\nu_{h2}$ are constant parameters of order $10^{-2}$. Finally,  

\begin{eqnarray}
Q(g) = \frac{\left|\frac{\partial^2 g}{\partial x^2}\right|}{|g|+\frac{1}{q}
	\left|\frac{\partial^2 g}{\partial x^2}\right|}
\end{eqnarray}

\noindent with $q$ the quench number, a constant of order ten, and $g$ a
function. An extra hyper-diffusive 
term is necessary for removing numerical errors coming from the ambipolar diffusion
where $\eta_{amb}$ is high and the current is low: 

\begin{eqnarray}
\frac{\partial {\bf B}}{\partial t} = ... + \frac{\partial}{\partial x} \left\{ \nu \Delta x 
\left[\frac{\nu_{a3} \nabla_{x}^{1}u_{Ax}}{|{\bf J \Delta x}|/|{\bf B}|} \right]
\frac{\partial {\bf B}}{\partial x} Q\left(\frac{\partial {\bf B}}{\partial x}\right)\right\}
\end{eqnarray}

\noindent where $\nu_{a3}$ is a constant parameter of order of $10^{-2}$ and 
$J \Delta x_x = (\partial Bz/\partial y ) \Delta y - (\partial By/\partial z ) \Delta z$. 
This hyper-diffusive splitting of the diffusive terms into local and global components 
makes it possible to run the code with a global diffusivity that is at least a 
factor 10 less than if the global term were the only one implemented in the code.
This implementation has been tested in \citet{Martinez-Sykora:2012uq}.

All the diffused magnetic energy coming from the  high-order artificial hyper-diffusion 
and the ambipolar diffusion is converted into thermal energy by Joule heating given 
by $Q_{Joule}={\bf E \cdot J}$  where the electric field ${\bf E}$ is calculated from the 
current ${\bf J}$.  

\section{Models and initial conditions}~\label{sec:models}

\begin{figure}
  \includegraphics[width=0.5\textwidth]{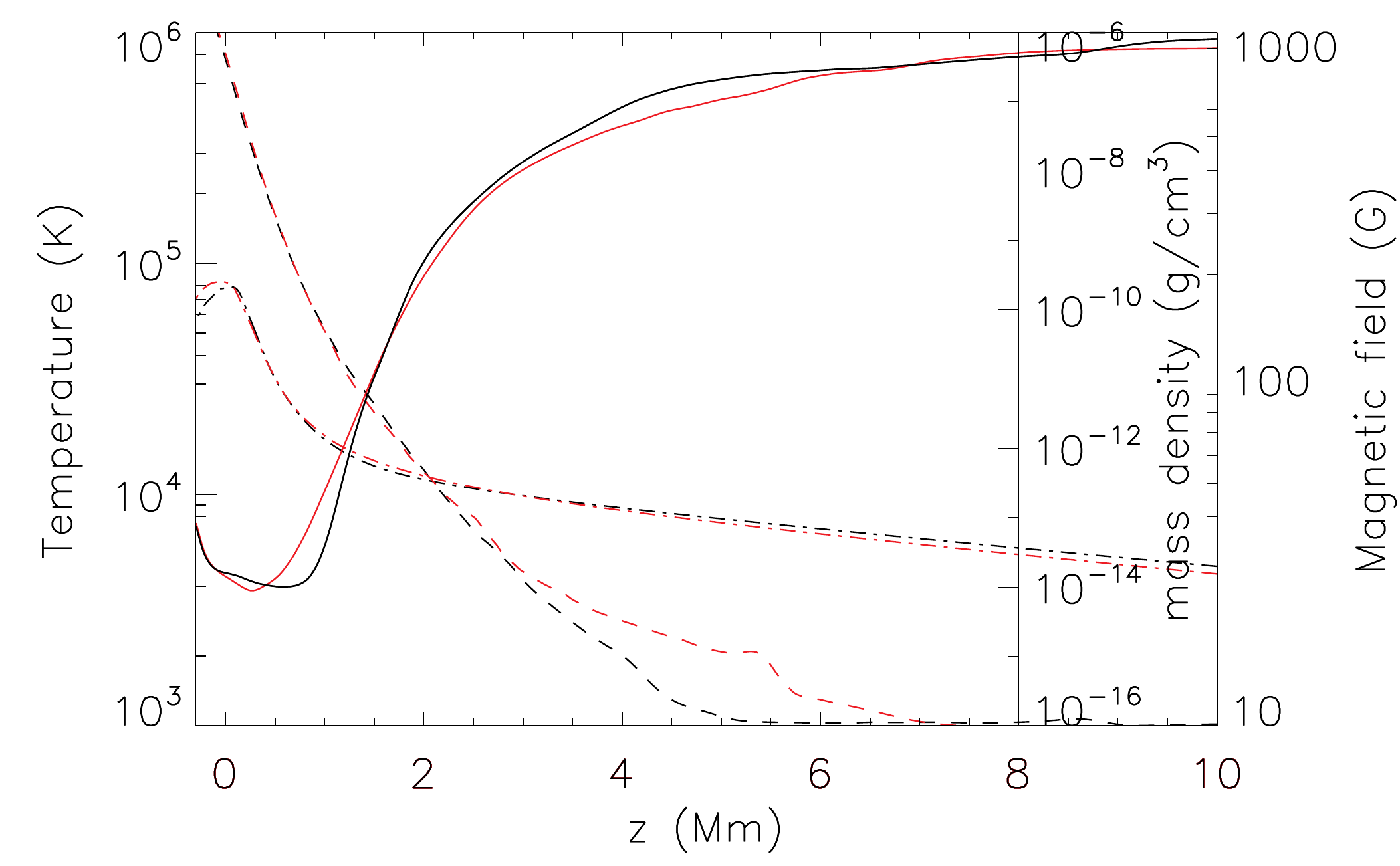} 
 \caption{\label{fig:1dstrat}  Horizontal and time averages for the temperature (solid line), unsigned 
 magnetic field (dash-dotted line), mass density (dashed line) for the non-GOL (black) and 
 GOL (red) simulations as a function of height integrated 
 between snapshots 300 and 370, i.e., over 11min.}
\end{figure}

The study presented here is focused on two different 2.5D models computed using 
the {\it Bifrost} code. Both 
models are calculated for a numerical domain that spans from the upper layers of the convection zone ($3$~Mm below the 
photosphere) to the corona ($40$~Mm above the photosphere). 
Convective motions perform work on the magnetic field and 
introduce magnetic field stresses in the corona. This energy is dissipated and creates 
the corona self-consistently as the energy deposited by Joule heating is spread 
through thermal conduction \citep{Gudiksen+Nordlund2002} and the temperature 
becomes on average about a million degrees 
(see solid lines in Figure~\ref{fig:1dstrat}).  Figure~\ref{fig:1dstrat} shows the
horizontal and time (integrated over 11 minutes) averages for the temperature, unsigned 
magnetic field, and mass density as a function of height. The horizontal domain 
spans $96$~Mm. The spatial resolution is uniform along the horizontal axis ($14$~km) 
and non-uniform in the vertical axis allowing smaller grid size where needed in 
certain  locations, i.e. the resolution from the photosphere (which has an effective temperature of 5800~K) to above the transition region ($z=7$~Mm) is $\sim12$~km while the grid spacing smoothly increases from the photosphere to the deeper layers of the convection zone and from $z=7$~Mm to greater heights up to $\sim 50$~km resolution.  

\begin{figure*}
  \includegraphics[width=0.99\textwidth]{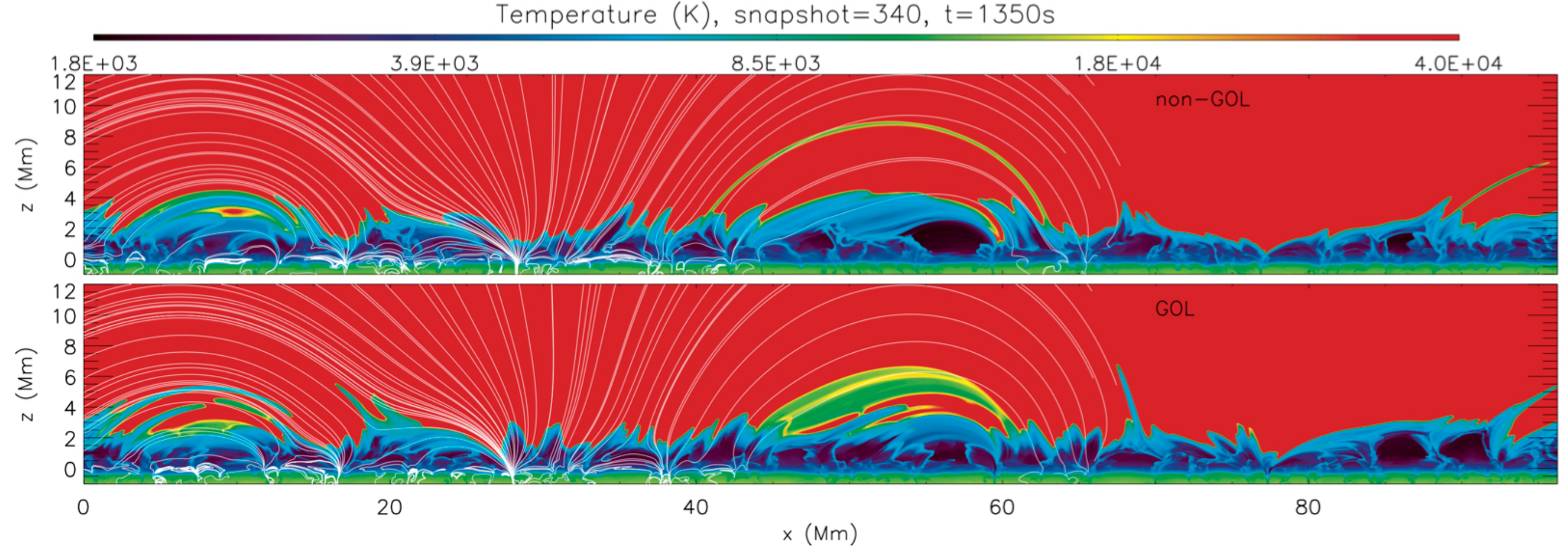}  
 \caption{\label{fig:2dtemp} Temperature maps for the non-GOL simulation (top panel) 
 and GOL simulation (bottom panel) reveal large differences in the thermal 
 properties. The temperature is shown in a logarithmic scale. Magnetic field lines are drawn  
 in the left hand side of the temperature map in order to show clearly the temperature map on the right hand side. }
\end{figure*}

The initial magnetic field has two medium size plage regions of opposite polarity that are connected and 
form loops that are up to $\sim50$~Mm long (see Figure~\ref{fig:2dtemp}). The mean unsigned field in the 
photosphere is $\sim190$~G (dash-dotted line in Figure~~\ref{fig:1dstrat}). 
The initial magnetic field is a potential field. First we run this setup
without ion-neutral interaction effects for roughly 35 minutes after transients have passed through 
the domain  (snapshot 200, t=0s). From this final instant, we run two simulations, one simulation 
incorporates the Generalized Ohm's Law, {\it i.e.}, includes the ion-neutral interaction effects 
(from now on we will refer to this effect and the model as GOL), and the other without (non-GOL). 
Each simulation was run for another $\sim 30$ minutes time after transients have disappeared.

\section{Results} \label{sec:res}

The ion-neutral interactions described above, and implemented through
the Generalized Ohm's Law, strongly influence the state of the simulated 
chromosphere. In the incoming sections, we will first focus on the
differences found in the statistical properties of the models with and
without GOL, {\it e.g}. the temperature, density, kinetic energy, and magnetic
field distribution. Following this, we will reveal the physical
mechanisms that are the root causes behind these statistical
differences. Finally, we will analyze specific chromospheric processes 
such as the expanding cold bubbles, jets, and reconnection, and
consider the role of the Generalized Ohm's Law in these.

\subsection{Thermodynamic properties}~\label{sec:termo}

The GOL simulation shows clear differences in
thermodynamic properties compared to the non-GOL simulation. This can be seen by
considering the Joint Probability Distribution Functions (JPDF) of the
density and temperature, using data integrated over 11 minutes and
shown in Figure 3. We call particular attention to cold, $T<4000$~K
($\log T<3.64$), tenuous, $\rho < 3.2\times 10^{-10}$g~cm$^{-3}$
($\log\rho<-9.5$) plasma. This is gas located in the wake of shocks,
``cold chromospheric bubbles'', which represent the lowest
temperature regions of the chromosphere \citep{Leenaarts:2011qy}, 
as long as there is no large-scale flux emergence 
\citep{paper1,Tortosa2009,Ortiz:2014wj}.
When ion-neutral interaction effects are not included we find that the temperature
in the wake of shocks is much lower than in the GOL model.
In fact, an {\it ad hoc} heating term is introduced in order to avoid plasma temperatures below 
$\sim 2000$~K, which are outside the validity range of the equation of state.
While this {\it ad hoc}
heating term is crucial for the non-GOL simulation, it is rarely
necessary in the GOL simulation since the cold chromospheric 
bubbles remain warm enough to prevent the {\it ad hoc} heating term
becoming active. 

The coolest regions in
the GOL simulation, for densities below $10^{-11}$ g~cm$^{-3}$, are several 
hundred of degrees hotter than in the non-GOL simulation (for which case artificial
and {\it ad hoc} heating dominate those locations). 
This results from being heated by the ambipolar 
diffusion when the cold bubbles 
reach low densities due to the expansion. 
In particular, the large number of neutrals and 
low ion-neutral collision frequencies in the cool bubbles cause significant dissipation of 
current perpendicular to the magnetic field (see Sections~\ref{sec:heating} and~\ref{sec:shocks}). 
This finding, {\it i.e.}, that the chromosphere remains relatively warm
in post-shock wakes, is compatible with observations of the molecular
lines of CO \citep{Penn:2011fk} which do not reveal 
absorption everywhere in the quiescent chromosphere. 
We speculate that the Joule heating contribution from ambipolar diffusion 
is large enough to prevent the low temperatures necessary for the formation of 
significant molecular line emission throughout the chromosphere. 
However, as a caveat, we also note that  
our simulations are representative for plage regions with rather
strong magnetic field. A firm conclusion on this issue will have to
wait for simulations that include magnetic field configurations that
are typical of quiet Sun.

\begin{figure}
  \includegraphics[width=0.49\textwidth]{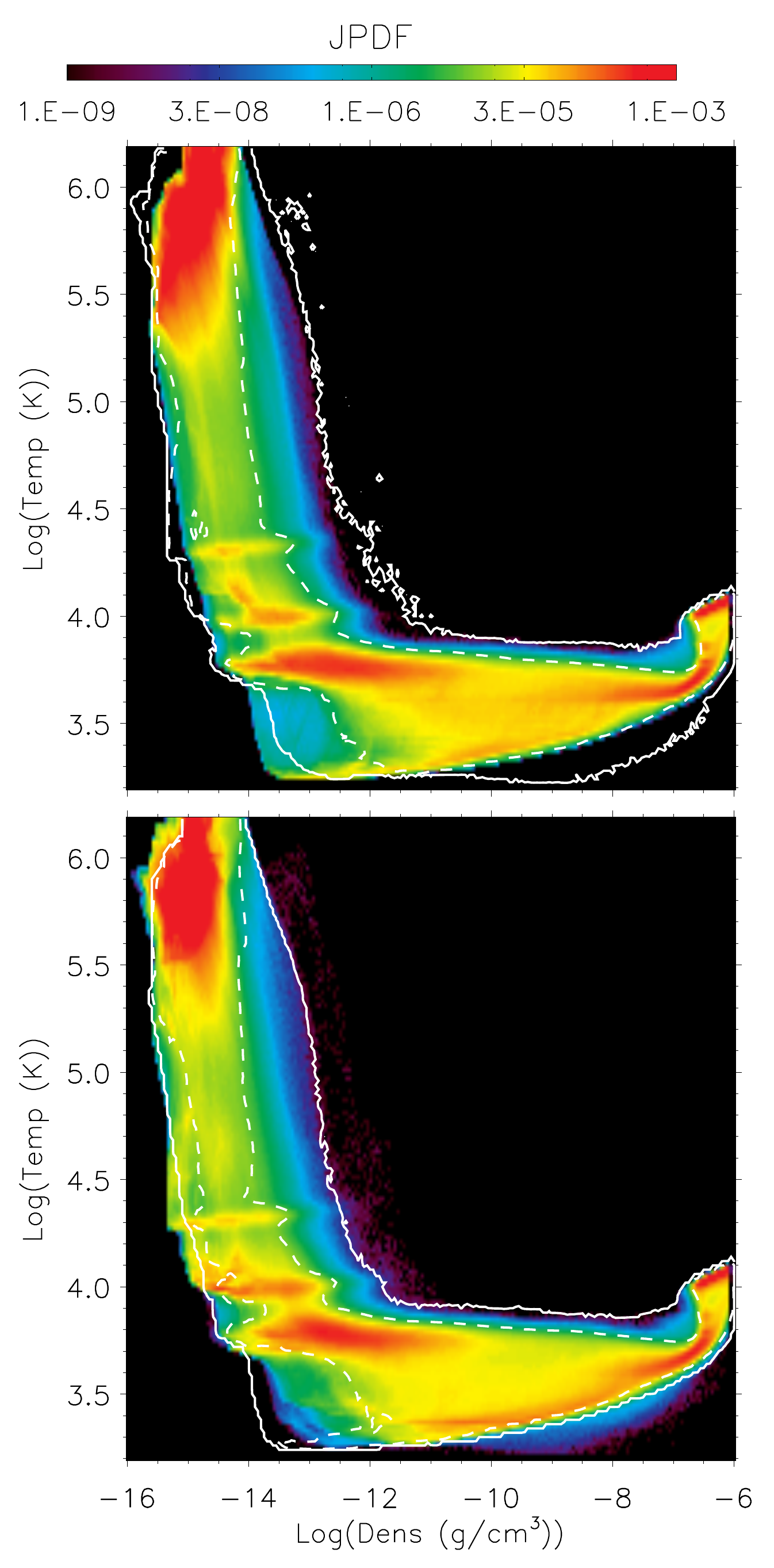}  
 \caption{\label{fig:histtgr} Joint Probability Density Function (JPDF) of temperature (vertical axis) and density (horizontal axis) over 11 min (between snapshots 300 and 370) for non-GOL (top) and GOL (bottom) simulations. The white contours correspond to the temperature and density regime of the whole simulation (solid) and at JPDF$=5\,10^{-5}$ (dashed) for non-GOL (bottom, whereas the color map corresponds to the simulation GOL) and GOL (top, whereas the color map corresponds to the simulation non-GOL) simulations in order to simplify the comparison. }
\end{figure}

In contrast to the tenuous cold bubbles, we find that in denser regions,  
close to the photosphere, the GOL simulation reaches cooler temperatures than the 
non-GOL simulation. This can be seen 
in the JPDF (Figure~\ref{fig:histtgr}) at temperatures below 4000~K
($\log T < 3.6$) and densities above $10^{-10}$~g~cm$^{-3}$
($\log \rho > -10$). This corresponds to a 
very small fraction  ($10^{-4}-10^{-5}$) of the upper photosphere and
lower chromosphere (dark blue in the color table). 
The blue area in the $-10 <\log \rho < -7$
region of the GOL JPDF (bottom panel) represents regions in which significant 
quantities of magnetic flux have accumulated in the photosphere. Close
to the upper photosphere, the ambipolar diffusion becomes large
enough to allow magnetic flux to expand 
through the photosphere and push material into the upper layers.
As this rising magnetic flux expands , it produces regions that are even
cooler than those found in the expanding cold bubbles driven by 
magneto-acoustic shocks \citep{paper1,Tortosa2009,Ortiz:2014wj}. 
This expansion of magnetic flux mediated by ambipolar diffusion 
does not happen uniformly over the full numerical domain but only 
sporadically, in a few locations where both the photospheric
magnetic flux and the ambipolar diffusion are large enough 
(see Sections~\ref{sec:jets} and~\ref{sec:chromjets}). 

We find that the upper chromosphere and transition region are more
extended in the GOL simulation  ($\sim1.3$ times) than the 
Non-GOL by comparing the JPDFs. There is an increase of plasma at temperatures between
5000~K ($\log T > 3.7$) and $10^5$~K ($\log T < 5$) with densities 
between $10^{-11}$~g~cm$^{-3}$ and $10^{-15}$~g~cm$^{-3}$. 
The spatio-temporal variation of the density is also greater in the upper chromosphere 
and TR: the density in the GOL simulation reaches both 
higher and lower densities. Similarly, the corona in the GOL simulation 
shows a wider range of densities (see the following sections for an explanation).  

\begin{figure*}
  \includegraphics[width=0.95\textwidth]{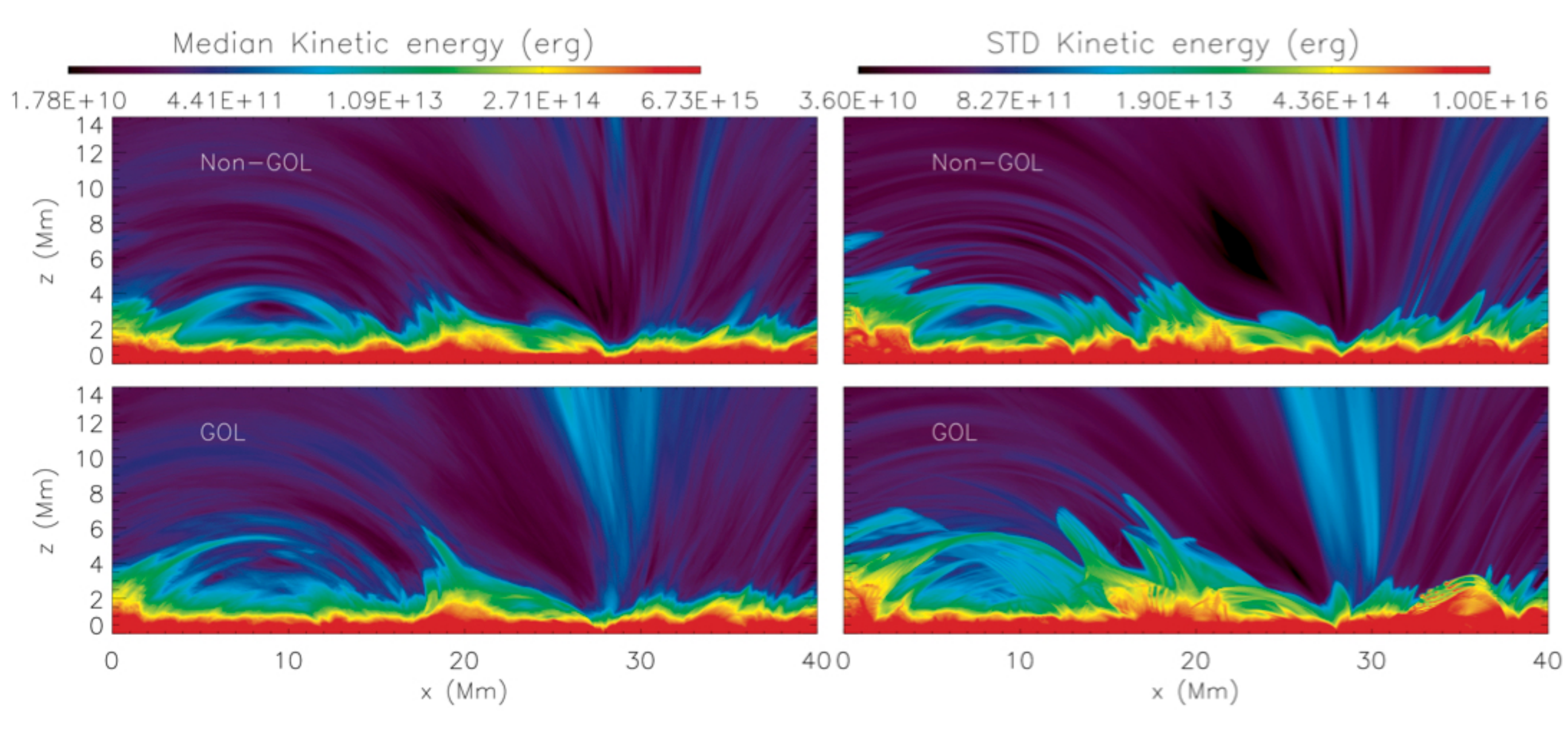} 
 \caption{\label{fig:kener} Median (left panels) and standard deviation (right panels) of the 
 kinetic energy (averaged over 11 minutes, snapshots=[300-370]) for the non-GOL (top panels)
 and GOL simulation (bottom panels), for $x=[0-40]$~Mm.}
\end{figure*}

Finally, the simulations also differ in terms of dynamics. Figure~\ref{fig:kener} shows the 
median (left panels) and standard deviation (right panels) of the kinetic 
energy (averaged over 10 minutes) for the non-GOL (top panel)
and GOL (bottom panel) simulation. 
The middle and upper chromosphere, as well as the TR, contain more  
kinetic energy and they show a larger range of values in the GOL simulation. This is due to the stronger
flows and other violent processes happening within the chromosphere (see Sections~\ref{sec:jets} and~\ref{sec:chromjets}). 
The enhancements of the kinetic energy and its range of values in the GOL simulation are 
located in regions where fast chromospheric jets or spicules occur or
where reconnection or strong currents are most likely to occur, {\it i.e.}, $x=[0-24]$~Mm, 
$x=[38-70]$~Mm, and $x=[80-96]$~Mm (see also Figure~\ref{fig:2dtemp},  Sections~\ref{sec:jets} and~\ref{sec:chromjets}). 
In this simulation, some regions with open field lines in the corona (x=30 Mm) shows enhancements of the kinetic 
energy which are caused by an artifact from the open boundary conditions at the top of the numerical domain. 
These can heat the open field lines under some conditions. The field lines associated with this artifact are 
localized to a small chromospheric region dominated by magneto-acoustic shocks. The artifact does not impact 
the lower chromosphere since the magneto-acoustic shocks are similar in nature as in other open field regions 
that are unaffected by the artifact.

\subsection{Energy distribution and field topology}~\label{sec:magnetic}

\begin{figure}
 \includegraphics[width=0.49\textwidth]{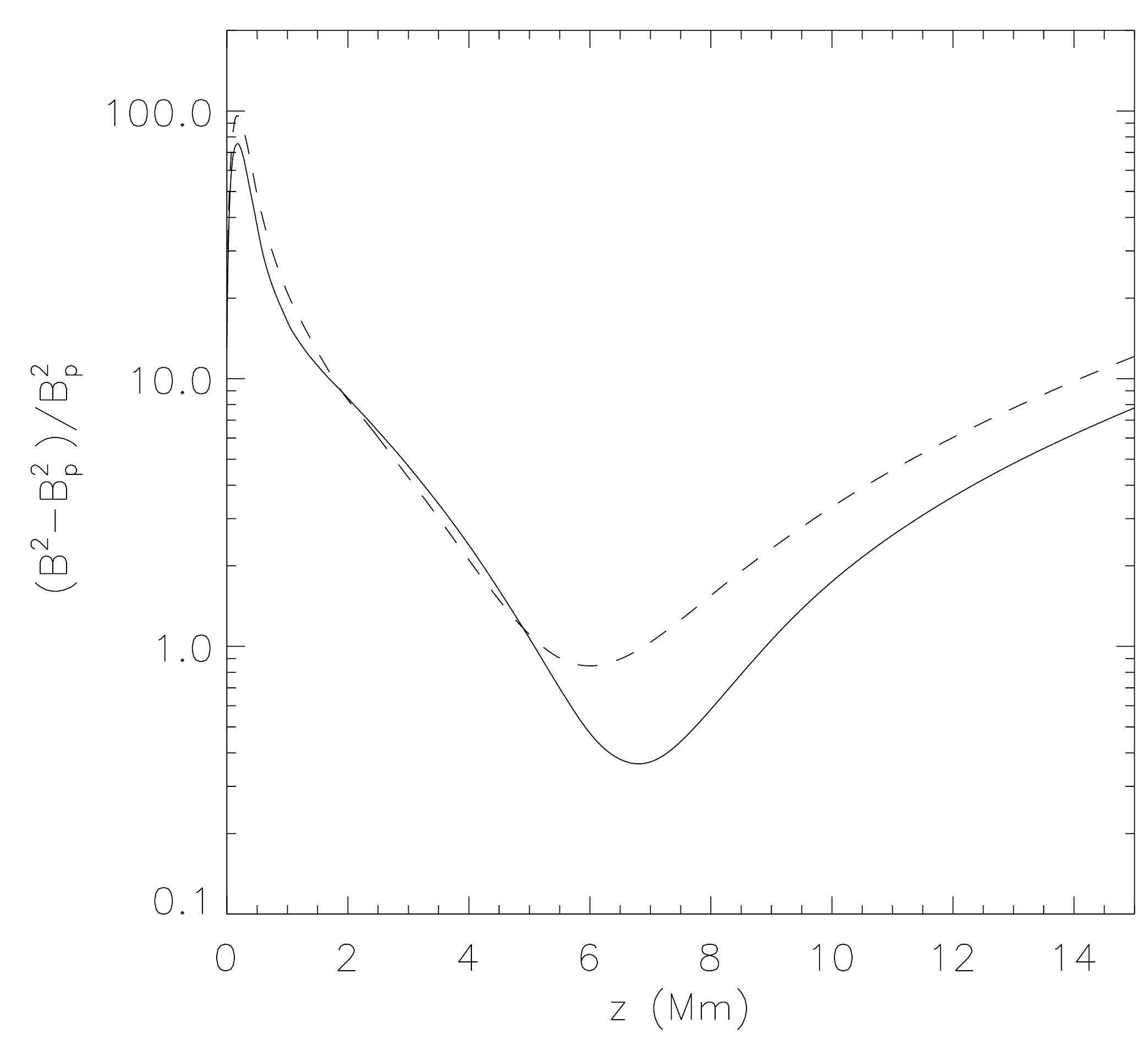}
  \caption{\label{fig:freeener} Magnetic free energy normalized by the 
magnetic energy of the potential field extrapolated from z=0~Mm for the GOL (dashed) and 
non-GOL simulations (solid) integrated over 11 minutes, (snapshots=[300-370]).}
\end{figure}

The magnetic field distribution and configuration is clearly different
in the two models. Figure~\ref{fig:freeener} shows, for both models, the magnetic free energy normalized by the magnetic energy of the potential field extrapolated from z=0 Mm. The magnetic free energy is calculated by subtracting the magnetic energy from a potential field extrapolation at $z=0$~Mm from the total magnetic energy and averaging this through the horizontal axis. The GOL simulation tends to accumulate slightly 
more magnetic free energy in the middle-upper chromosphere and TR ($z=[2,5]$~Mm) but less magnetic free energy 
in the corona.  In other words, the ion-neutral effects appear to prevent 
free energy from reaching the corona, instead accumulating
non-potential field in the chromosphere, where it is dissipated. The
dissipation is in part violent and leads to the
increased thermal and kinetic energy we find in the GOL simulation (in certain locations with timescales ($e/Q_{j_{amb}}$) of a few seconds or even shorter). 
There are several reasons for this (see also 
the following sections):

\begin{itemize}
\item Our simulations do not include imposed flux emergence through
  the bottom boundary; however, the convective motions
  cause horizontal magnetic fields to accumulate in the sub-adiabatic photosphere (see 
Section~\ref{sec:chromjets}). Strong concentrations of magnetic flux
can lead to a reduced gas pressure which 
may facilitate the  onset of the Rayleigh-Taylor instability (RT, which is 
caused by denser plasma overlying less dense plasma). However, the photosphere is 
sub-adiabatic, and therefore has strong stabilizing properties
\citep{Acheson:1979lr,archontis2004}.  
As suggested by \citet{Leake:2006kx} and \citet{Leake:2013dq} the ion-neutral inter- action facilitates to diffuse the magnetic field through the photosphere 
\citep{Acheson:1979lr} allowing more magnetic flux to penetrate into the chromosphere than what occurs in the non-GOL simulation. 
\item However, ambipolar diffusion does not always facilitate the diffusion of magnetic field into the chromosphere. One way to visualize this is to consider Equation~\ref{eq:faradtot3} in which the ambipolar term in the induction equation is written as an advection term. The ambipolar velocity diffuses magnetic field into the chromosphere if it is oriented upwards, and remove chromospheric field when it is directed downwards. The latter occurs when the (horizontal) magnetic field strength increases with height which will lead to downwards ambipolar velocities. 
\item The ambipolar diffusion can not only diffuse magnetic field but 
also concentrate magnetic flux, which will help to 
promote RT instability due to buoyancy. This can be explained as follows: In cases with a
horizontal flux tube with twist, in the upper part of the tube, the orientation of the 
current and magnetic field lines, following the right hand rule, will always lead to an ambipolar velocity 
pointing towards the convection zone. On the other hand, in the 
lower part of the tube, the current points in the same direction as in the upper part of the tube 
while the magnetic field lines are oriented in the opposite direction,
the ambipolar velocity will push the magnetic field lines towards the 
corona (see Sections~\ref{sec:flux}, \ref{sec:jets}
and~\ref{sec:chromjets}). In other words,  
ambipolar diffusion tends to compress a horizontal twisted flux tube. 
This leads to a decrease of the gas pressure, and thus an increased buoyancy. 
\item We find more magnetic free energy in the middle-upper
  chromosphere and TR due to the excess magnetic flux
  that expands from below as a result of ambipolar diffusion.
\item Ambipolar diffusion tends to accumulate current
 in narrow layers. This leads to faster reconnection 
 rates and larger magnetic tension in the chromosphere.
\item Since the chromosphere transforms magnetic energy into thermal and kinetic 
energy more efficiently due to ambipolar diffusion, the corona in the
GOL simulations contains less magnetic free energy than the non-GOL simulation. 
\end{itemize}

\begin{figure}
  \includegraphics[width=0.49\textwidth]{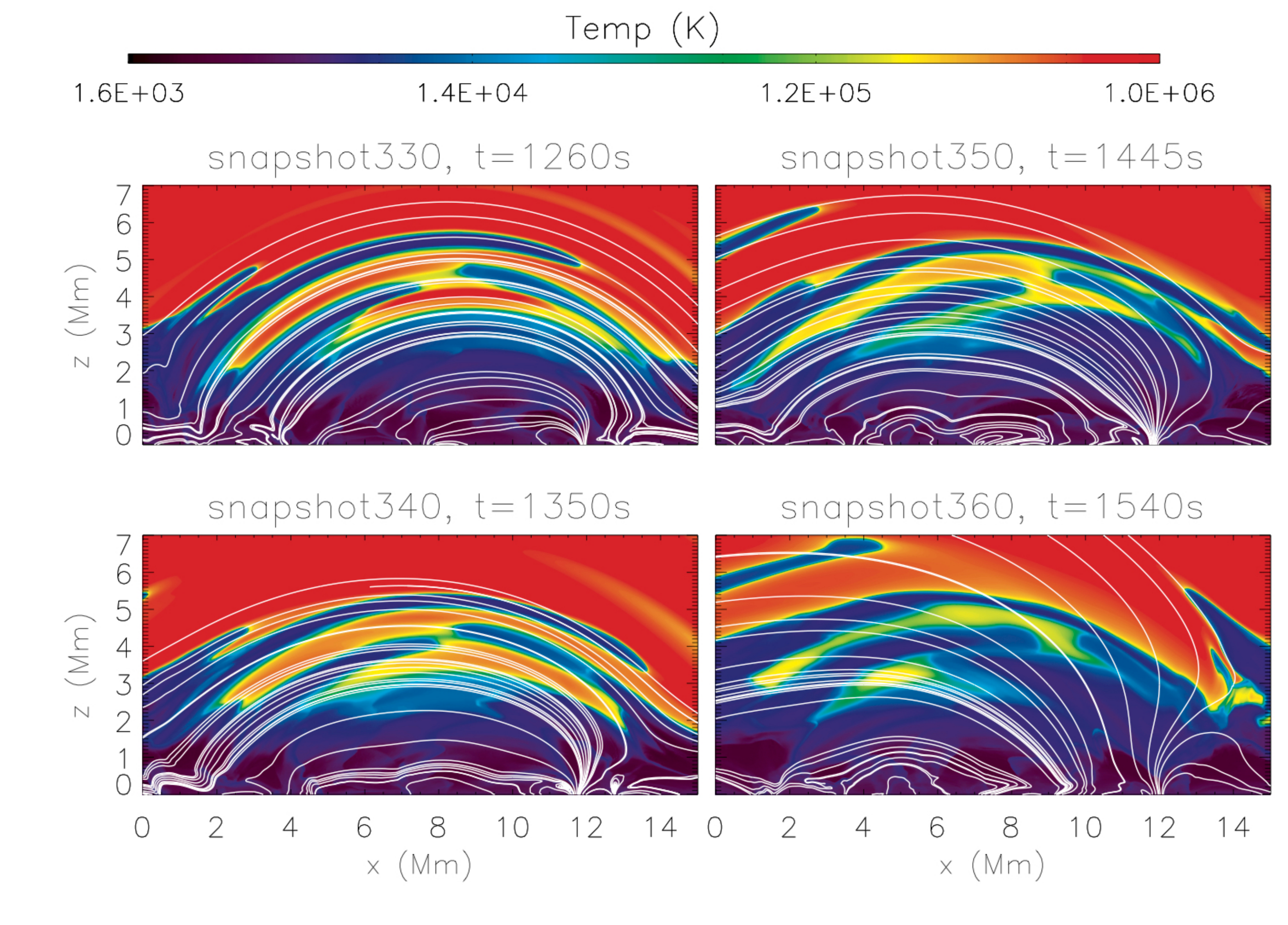}
  \caption{\label{fig:fields} Temperature, in logarithmic scale, for the GOL simulation at 
  t=1260s (top-left), 1350s (bottom-left), 1445s (top-right) and 1540s (bottom-right) with
  magnetic field lines shown in white. In the beginning (left panels) the thermodynamic structures
  are aligned with the magnetic field, whereas at later times (right panels) the magnetic connectivity
  has changed and the alignment is poor (see corresponding Movie 1). }
\end{figure}

We also find that as a result of the ambipolar ``velocity'' 
thermodynamic structures may decouple from the magnetic field under
certain circumstances. When ambipolar diffusion,
magnetic field strength and the current perpendicular to the magnetic 
field are high enough, and the time-scales of the thermodynamic processes are longer than 
the ambipolar processes, the thermodynamic structuring does not necessarily follow the magnetic field direction. One example of this is shown in Figure~\ref{fig:fields}. Before the 
ambipolar velocity decouples the magnetic field structures from the thermal properties, one 
can see that the magnetic field collimates the jets and that the transition region loops
also follow the magnetic field direction (left panels). Later in time,
the magnetic field lines rooted at $x=12$ Mm 
{\it move} from right to left, and in the lower photosphere the magnetic field
connectivity changes so that the magnetic field lines
connect to different photospheric locations. The evolution of
the magnetic field connectivity is faster than or 
of the same order as
the thermodynamic evolution. As a result, the magnetic field no longer
aligns well with the thermodynamic 
structures in the upper chromosphere and TR. For
example, the magnetic field lines 
cross the jet diagonally on the left side of the bottom-right panel. 
This will impact the evolution of the features (see the drift towards the left of 
the left magnetic field footpoint in the GOL simulations in the Movie 1). \citet{Martinez-Sykora:2016qf} describes in more detail the misalignment 
of the magnetic field lines with thermal structures. 
 These results may provide an explanation 
for the puzzling observations of \citet{de-la-Cruz-Rodriguez:2011qd} and \citet{Asensio-Ramos:2017rp} 
where fibrils appear to not necessarily follow 
the magnetic field structures.

\begin{figure}
  \includegraphics[width=0.49\textwidth]{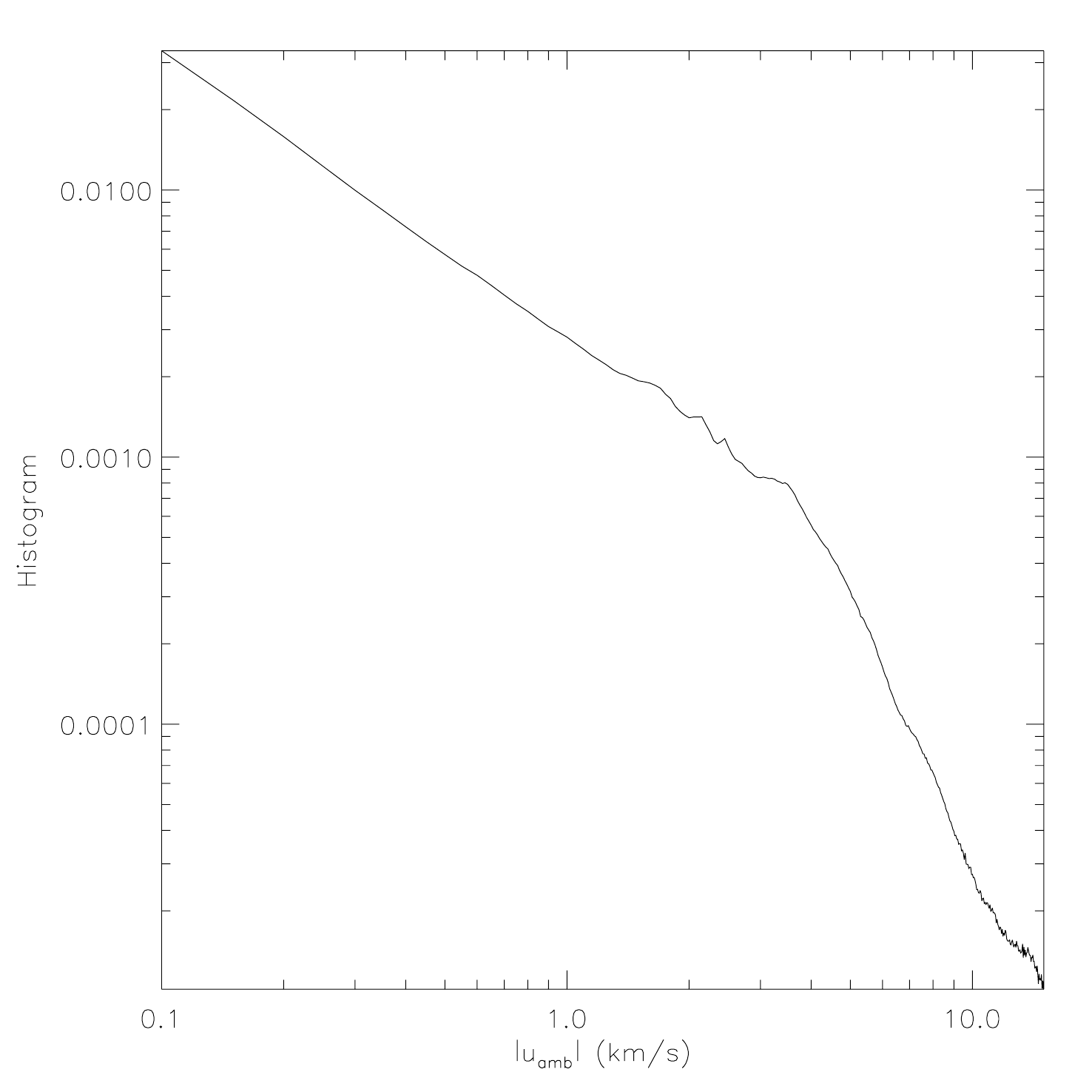}
  \caption{\label{fig:histoua} Histogram of the absolute value of the ambipolar velocity ($|u_{amb}|$), integrated over 11 minutes (snapshots$=[300-370]$) reveals a two step power-law (with a slope of $-0.07$ at low ambipolar velocities and $-0.2$ at high ambipolar velocities).}
\end{figure}

The term in the induction equation that leads to the misalignment is
that of the ambipolar velocity. For the plage-like magnetic field
configuration that we simulate here, a histogram of the ambipolar
velocity shows a power-law like behavior in two distinct regions (see
Figure~\ref{fig:histoua}). The power law slope 
for the lower values of the ambipolar velocity is due to the processes 
happening in the lower chromosphere, where the plasma is denser and the ambipolar 
diffusion is smaller. This region is dominated by the shocks driven by the 
convection zone and the photospheric overshooting. 
Here the ambipolar velocity distribution is smoother and the
locations where the ambipolar velocity is significant are more spatially extended 
than in the upper chromosphere (where they are more concentrated spatially).
In contrast, in the upper chromosphere, the ambipolar velocity is larger and concentrated
in narrower regions along the spicules or loops. 

\subsection{Heating properties}~\label{sec:heating}

The most obvious difference between these two simulations and previous
2D simulations is that the convective motions below and up to the photosphere stress the magnetic
field sufficiently to self-consistently maintain a hot 
corona (Figure~\ref{fig:1dstrat}). Previous 2D MHD simulations 
\citep{Leenaarts:2011qy,Heggland:2011kx,Iijima:2015fk,Nobrega-Siverio:2016qf}
have required a hot plate at the upper boundary in order to produce a hot corona. 
Previously it was only when computing 3D models that self-consistently heated coronae arise 
\citep[\eg][]{Gudiksen+Nordlund2002,Hansteen:2010uq,Martinez-Sykora:2011oq,Hansteen:2015qv,Carlsson:2016rt}. 
In the current simulations, we find that despite the 2D limitation, the large scale magnetic field configuration in the current simulation leads to a self-consistently heated corona. This comes about 
as a result of the large variety of processes that occur simultaneously within the simulated
domain.

In fact, the Joule heating in both of these two simulations is not only greater
than in previous 2D models, but also extends over a wider range of
heights. In previous simulations, most of the
heating per particle was strongly confined to the TR.  
This seems to be in accordance with
  \citet{Hansteen:2015qv} who suggested that large scale 
connectivity leads to a larger scale height for the heating per
particle. Our configuration has a smaller decay of Joule heating as a function of height than those smaller scale simulations in  \citet{Hansteen:2015qv}.

\begin{figure*}
  \includegraphics[width=0.95\textwidth]{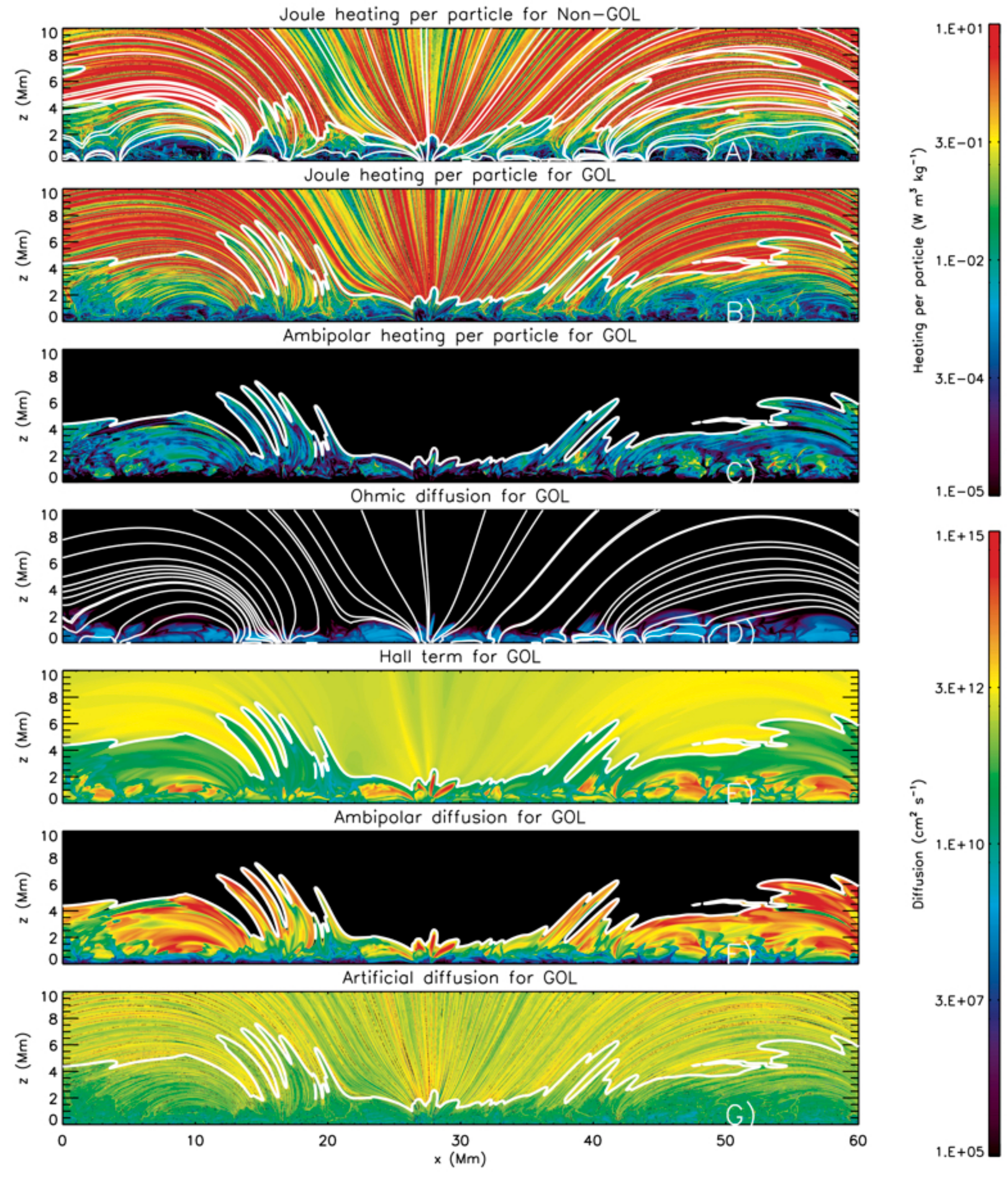} 
 \caption{\label{fig:heat} From top to bottom: maps of the Joule heating from the artificial 
 diffusion for the non-GOL (panel A) and GOL simulation (panel B) and from the 
 ambipolar diffusion (panel C), and ohmic diffusion (panel D), Hall term (panel E), ambipolar  (panel F) and 
 numerical diffusion  (panel G) at t=800s (snapshot=280).
 Panels A and D show magnetic field lines in white and the other panels show the 
 transition from the chromosphere to the corona as white contours (contour of $T=10^5$~K).
These maps reveal the importance of Joule heating from the ambipolar diffusion in 
 the cold bubbles and upper chromosphere. The color schemes are in logarithmic scale.}
\end{figure*}

In order to interpret the various heating mechanisms in these models, it is important to show that regions in which ambipolar diffusion dominates are well 
resolved:
\begin{itemize}
\item The artificial and ambipolar diffusion act differently,
  depending on local magnetic field and thermodynamic conditions (\eg\ compare the 
ambipolar heating in panel C with artificial Joule heating in panels A and B of Figure~\ref{fig:heat}). 
These two different heating mechanisms are dominant
in different regions in the chromosphere (panels G and F 
in Figure~\ref{fig:heat}). 
\item Our 2.5D simulation reveals that ambipolar diffusion is very important in the chromosphere
(panel F in Figure~\ref{fig:heat}). 
In extended regions in the chromosphere, ambipolar diffusion is much larger than the artificial 
diffusion (compare panels F and G in Figure~\ref{fig:heat}). 
\item The GOL and non-GOL models differ in their magneto-thermo-dynamic 
properties as detailed in previous sections (Figure~\ref{fig:histtgr}). 
\end{itemize}
The ambipolar diffusion strongly depends on the thermal properties of the 
chromosphere and TR. This behavior cannot be captured with 
1D semi-empirical models because those do not capture the dynamics of the
chromosphere, {\it e.g.} shocks, or the significant
horizontal spatial structuring. The large variations in ambipolar diffusion 
have significant consequences for a variety of physical processes
(Section~\ref{sec:shocks}-\ref{sec:chromjets}). 
This large variability of the ambipolar diffusion is due to the 
significant changes in the ion-neutral collision frequency 
which depends on the neutral density, the ionization state and the
temperature. 

In terms of absolute values, the ambipolar diffusion term is the
largest when comparing to the ohmic diffusion, Hall term, and
artificial diffusion. This is shown in panels D, E, F, and G in 
Figure~\ref{fig:heat}. The ohmic diffusion (panel D) is larger in the photosphere and 
chromosphere than in the corona. The Hall term (panel E) is important in the corona and in the 
coolest areas in the chromosphere \citep[see ][ for details]{Martinez-Sykora:2012uq}.Despite this, the Hall term in our 2.5D numerical model did not reveal any appreciable impact on the simulated atmosphere. Most likely one may need to expand the simulation to 3 dimensions, or include greater spatial resolution (e.g., to resolve Whistler waves \citep[e.g.,][]{Huba:2003fk}
 
The Joule heating caused by ambipolar diffusion 
in the GOL simulation (panel C of Figure~\ref{fig:heat}) 
is mostly  localized in two types of regions: 1) The cold chromospheric 
bubbles produced by the rarefaction in the wake of shocks that pass through the 
chromosphere. This heating is larger in the upper regions of the cold expanding bubbles 
where the density, temperatures and ion-neutral collision frequency are lower. 
The ambipolar heating is not very important in denser regions, {\it i.e.}, close to the photosphere. 
2) The ambipolar diffusion is also very important in the upper chromosphere. 
There we find large spicule-like extrusions of the TR 
that roughly align with the inclined magnetic field (regions 
between $x=[13-23]$~Mm and  $x=[37-43]$~Mm). These are heated by 
ambipolar diffusion. Similarly, the upper chromosphere in regions with highly 
inclined (almost horizontal) magnetic field are also heated by ambipolar 
heating (regions between $x=[0-10]$~Mm and  $x=[45-60]$~Mm).

It is interesting to see that in the cold bubbles in the lower
atmosphere, where the ratio of neutral to ionized particles is very
large \citep[see also][]{Martinez-Sykora:2012uq}, 
the Joule heating from the artificial diffusion (panel B) is smaller 
than both the same heating term in the non-GOL simulation (panel A) and 
the Joule heating from the ambipolar diffusion 
(panel C). However, in the upper 
chromosphere, TR and lower corona
the Joule heating from the artificial diffusion (panel B) is larger  
than in non-GOL simulation (panel A). 
This is caused by a combination of two effects: 1) the 
ambipolar diffusion allows more magnetic flux to reach into the 
chromosphere which through interaction with the ambient, pre-existing
field leads to more heating (see Sections~\ref{sec:flux} and ~\ref{sec:chromjets}), 
2) the ambipolar diffusion concentrates 
the electrical current which is dissipated by both ambipolar diffusion and artificial 
diffusion.

\subsection{Magnetic flux transport}~\label{sec:flux}

\begin{figure}
  \includegraphics[width=0.5\textwidth]{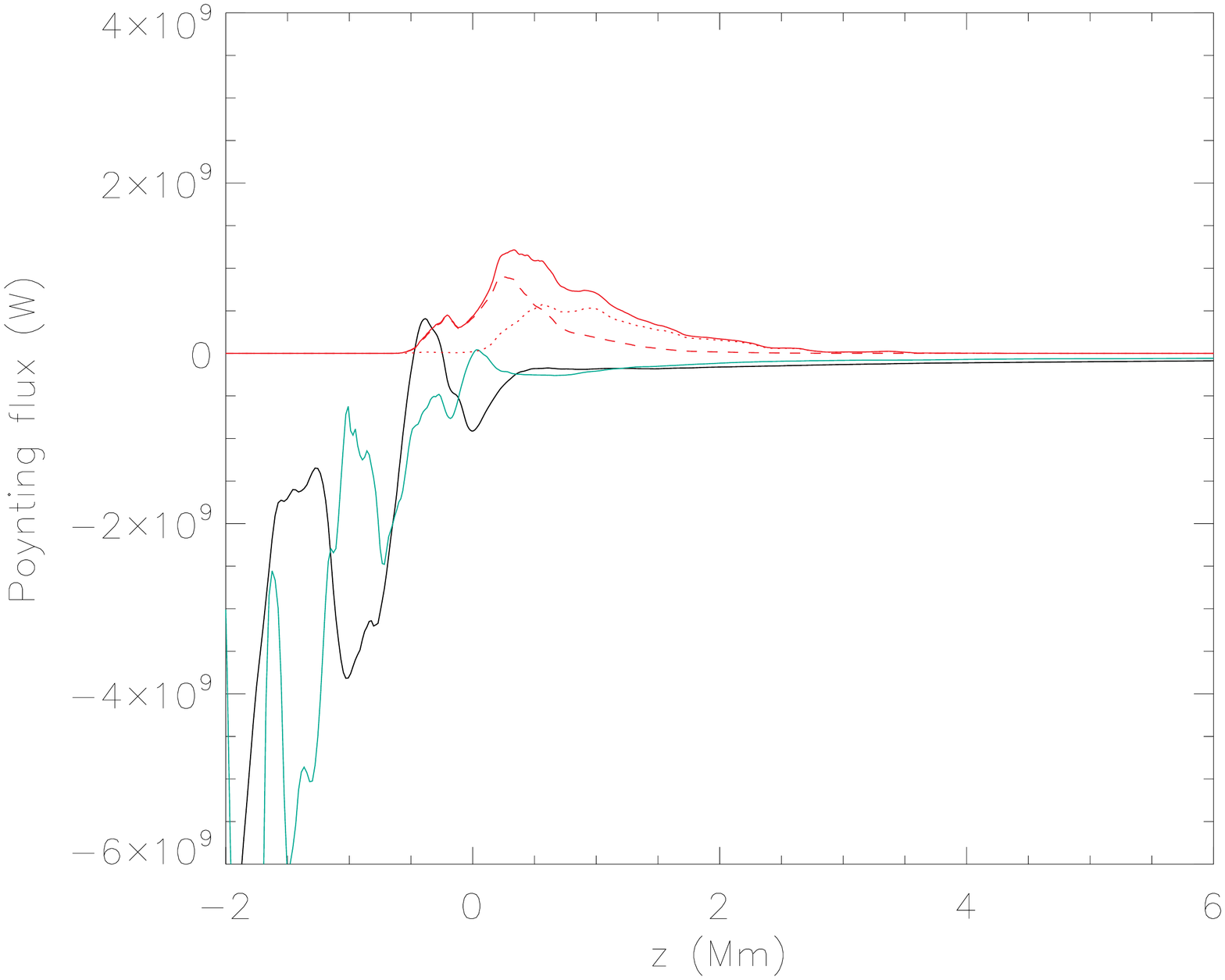} 
 \caption{\label{fig:flux} Total vertical Poynting flux of the advection term 
 for the non-GOL (black) and the GOL simulations (green) as a function of height. All values 
 are averaged in time (over 1.5 minutes,  snapshots=[360-370]) and in the horizontal direction. The 
 vertical Poynting flux due to the ambipolar diffusion in  
 the GOL simulation (red) is shown as a function of height. The dashed line is for horizontal motions
  (${P_z}_h$), dotted line is for vertical motions (${P_z}_z$) and solid is the sum of both. }
\end{figure}

Sporadically the ambipolar diffusion releases 
photospheric magnetic field into the chromosphere.
Since the Hall and ambipolar terms can be rewritten as in Equation~\ref{eq:faradtot3} they 
may be interpreted as a Poynting flux where the velocities are the Hall and ambipolar ``velocities'', 
respectively. Therefore, one can calculate the vertical 
Poynting flux due to vertical and horizontal components of the ambipolar velocity:

\begin{eqnarray} 
{\bf P_z}_z \equiv {\bf {u_A}_z}(B_x^2+B_y^2) \\
{\bf P_z}_h \equiv {\bf B_z}(B_x {u_A}_x+B_y {u_A}_y)
\end{eqnarray}

\noindent which are shown in Figure~\ref{fig:flux} as red dotted and red dashed 
lines respectively. The sum of both is shown with the solid red line.
This figure also shows the vertical Poynting 
flux (only the contribution from the advection) 
for the non-GOL (black) and GOL simulations (green) as a function of height. The values
are integrated over a time period of two minutes.

Ambipolar diffusion can facilitate the transport of magnetic field from the photosphere
into the chromosphere by $\sim10^9$~W. This does not occur universally since the effect is concentrated 
in a few locations or events instead of being uniformly distributed over the upper-photosphere (See the example in Section~\ref{sec:jets}). The Poynting 
flux due to the ambipolar ``velocity" is important in the proximity of the photosphere as 
can be appreciated from the integrated ${\bf P_z}_z$ and ${\bf P_z}_h$ as a function of 
height in dashed and dotted red lines.
As a result, in some colder than average locations 
additional magnetic flux is carried into the middle chromosphere. 
At these locations the magnetic field expands quite rapidly, and this expanding magnetic field interacts 
with the ambient magnetic field (see Section~\ref{sec:jets} 
and~\ref{sec:chromjets}). This is similar to what \citet{Leake2006} found in their
2D simulations of flux emergence, where partial ionization effects were seen
to facilitate the expansion of the magnetic field into the upper solar
atmosphere and corona. However, there are two major
differences between our and their simulations: Our simulation do not include any imposed 
flux emergence, and the ambipolar diffusion is critical for the horizontal field to 
penetrate into the chromosphere. In contrast, in \citet{Leake2006} emerging flux crosses the 
photosphere due to the strong buoyancy, while ambipolar diffusion did not 
play a large role. In their case ambipolar diffusion was not strong in
the upper-photosphere as they used a 1D semi-empirical model (which does not
capture the extremely low temperatures in the wake of shocks) to describe the
upper-photosphere. When the temperature and ionization dependence of the ambipolar diffusion are taken
into account, as in our model, large variations of the ambipolar
diffusion (Figure~\ref{fig:heat}) occur at photospheric heights. 

The rapid expansion due to the emergence of flux through the
photosphere produces cool voids. This occurs in  
denser regions (blue regions in GOL simulation for $-10<\log \rho <-7$, Fig.~\ref{fig:histtgr}) than the cold bubbles that arise as a result of 
rarefaction behind magneto-acoustic shocks ($\log \rho < -10$).
These types of voids are quite rare because of the following reasons: 1) the ambipolar diffusion 
is rarely significant enough in the vicinity of the photosphere to
trigger the instability that leads to flux emergence, 2) we do not include 
any explicit flux emergence.  
As a result of the increased flux emergence from the ambipolar diffusion, the magnetic field strength in the 
photosphere is slightly reduced compared to the non-GOL simulation.

\subsection{Detailed description of the various physical processes}~\label{sec:processes}

We will now describe some of the dominant physical processes in these simulations 
and how they differ between the two simulations. 
Our large scale models reveal, for the first time in this type of 
models, that the magnetic field configuration and connectivity plays a
key role in the nature of the dominant magneto-thermo-dynamic
processes in the various regions of the domain. We can distinguish several
different types of regions in Figure~\ref{fig:2dtemp}: 1) open field
lines in the two (opposite polarity) plage regions (at
$x=[25-35]$~Mm and at $x\sim[60-80]$~Mm), 
dominated by magneto-acoustic shocks 
(dynamic fibrils); 2) inclined magnetic field lines
that penetrate into the corona and connect both plage regions next to the open field lines ($x=[19-25]$~Mm, $x=[35-50]$~Mm, 
$x=[67-70]$ and $x=[80-96]$~Mm), dominated by taller spicules and
jets. Note the overlap between various regions due to the complexity 
of the connectivity; 3) regions where the chromospheric magnetic field lines 
are highly inclined or almost horizontal and never reach the corona 
($x=[2-18]$~Mm and $x=[50-60]$~Mm).  The latter are located in regions between 
the two polarities. The physical processes 
that we will describe in detail are the magneto-acoustic shocks, chromospheric
jets, and the reconnection processes that occur in 
regions with highly inclined magnetic field. The different
spatio-temporal evolution of these processes between the two
simulations is ultimately the cause for the different thermodynamic 
stratifications that we described in the previous sections. 

\subsubsection{Expanding ``cold" rarefraction bubbles}~\label{sec:shocks}

\begin{figure*}
  \includegraphics[width=0.99\textwidth]{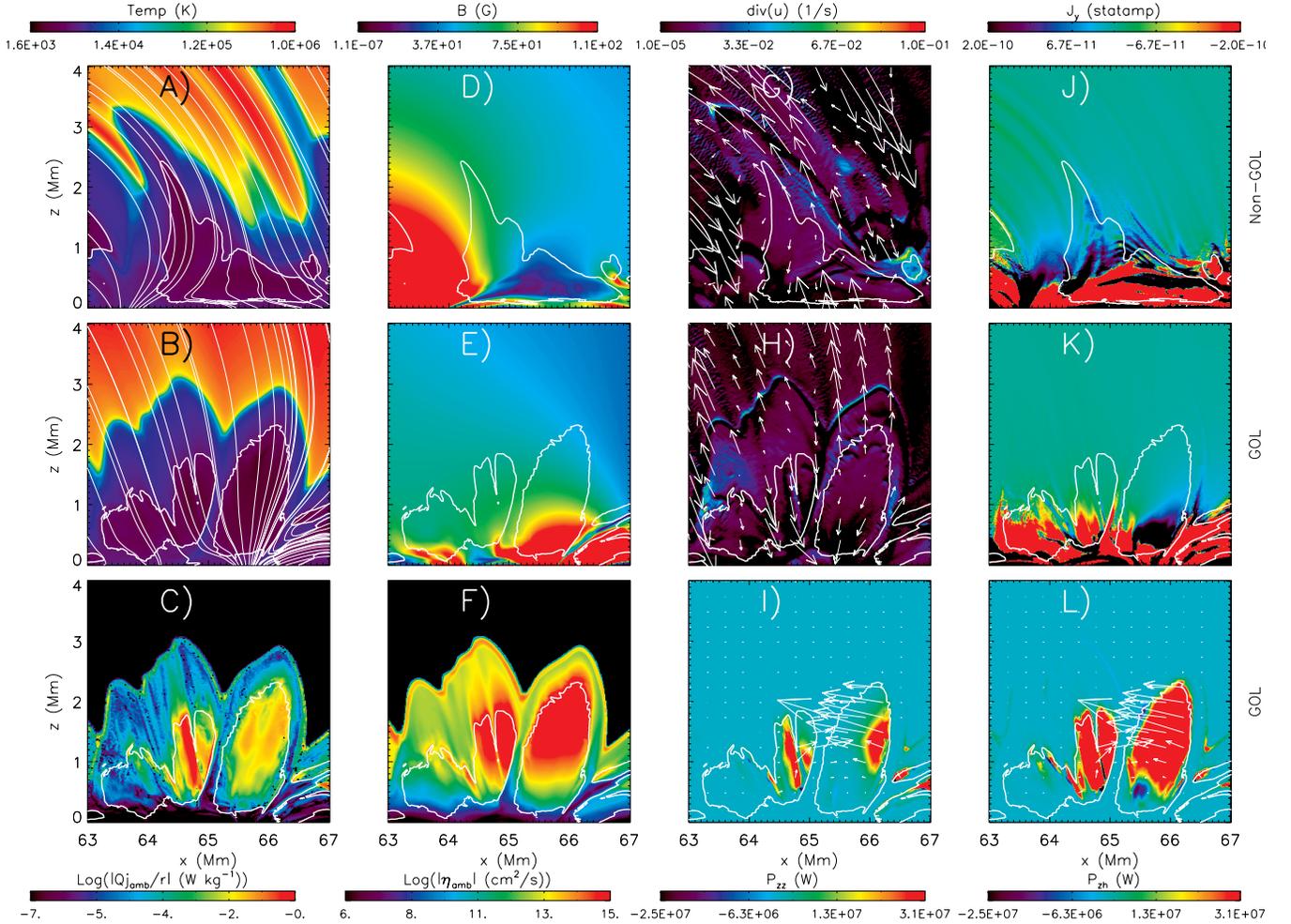}
\caption{\label{fig:shockmaps} The top row shows maps of temperature 
  (with magnetic field lines in white, panel A), density (panel D), divergence of the
  velocity (with velocity field as white arrows, panel G), electric current density perpendicular to the 
plane (panel J) for the non-GOL simulations at
	t=1240s, snapshot=328. Middle row is the same as
the top row, but for the GOL simulation (panels B, E, H and K) at
t=1350s, snapshot=340. The
bottom row shows, for the GOL simulation: Joule heating from 
the ambipolar diffusion (panel C), ambipolar diffusion (panel F), vertical Poynting 
flux due to the vertical ambipolar velocity (panel I) and the vertical
Poynting flux due to the horizontal ambipolar velocity (panel L). 
The ambipolar velocity field is shown with white arrows in panels I and L. 
This region is a representative for magneto-acoustic shocks that go though the 
region between $x=[25,45]$~Mm and $x=[70,80]$~Mm which does not show any dramatic reconnection. The white 
thick contours correspond to a $4000$~K temperature.}
\end{figure*}

As a result of the large ambipolar diffusion in the cold bubbles 
in the wake of strong acoustic shocks, we find differences in several
aspects compared to the non-GOL simulation. This is true not only for the
thermal properties, but also for the magneto-dynamic
properties of these bubbles. Ambipolar diffusion reduces, and in many cases removes, 
the current density in the inner upper part of the bubble  through
dissipation into thermal energy. The cold bubbles in the non-GOL 
simulation typically show significant current throughout (panel J,
Figure~\ref{fig:shockmaps}). This is not the case in the GOL
simulation, where the current density is typically removed due to
ambipolar diffusion, especially in the upper part of the cold bubbles
(panel K). As a result, the magnetic field in the
GOL simulation tends to be nearer to potential and more uniformly distributed in the cold chromospheric bubbles, despite their strong 
expansion. Since the magnetic field is less uniformly distributed in the non-GOL 
simulation, we find the lowest magnetic field strengths in the
cold bubbles in the non-GOL simulation ({\it e.g.}, compare the bubbles 
at $x\sim65$~Mm in panel D of Figure~\ref{fig:shockmaps} with the ones $x=65-66$~Mm
in panel E). In the non-GOL simulation we also see that in regions with
predominantly vertical fields the magnetic field strength shows 
stronger variations than in the GOL simulation (not shown in the figure).

Note that not all cold bubbles are characterized by significant 
ambipolar diffusion, as this depends on the bubbles' density, 
ion-neutral collision frequency, and magnetic field
strength. If the ambipolar diffusion is strong we find that at later 
stages of a bubble's evolution the current density
is entirely removed from the upper part of the bubble and the
remaining current tends to be located at the sides of the bubble 
(see at $x=66.5$~Mm in panel K). The current density is concentrated
there as a result of the interaction between the magnetic field in the expanding bubble, which suffers
from the strong ambipolar diffusion, and its surroundings (see 
the ambipolar diffusion in panel F of Figure~\ref{fig:shockmaps}). 

The expanding cold bubbles found in the GOL simulation are warmer 
(Figure~\ref{fig:histtgr}) as a result of the increased magnetic field
dissipation caused by ambipolar diffusion (see panels C, F in Figure~\ref{fig:shockmaps}).

It is also illustrative to consider the ambipolar velocity field which is shown in panels I and L with 
white vectors for the GOL simulation.  

In the spicules, which are mostly vertical, the ambipolar velocity is almost horizontal because the 
magnetic field is vertical and the current is perpendicular to the plane of the simulation. 
Therefore, the ambipolar advection of the magnetic field lines leads to the 
leftward drift of the field lines of the leftmost plage region in the GOL simulation,
which explains the horizontal drift of the magnetic 
field shown in Movie 1, 
Figure~\ref{fig:fields} and Section~\ref{sec:magnetic}.

\subsubsection{Chromospheric jets}~\label{sec:jets}

We find several jets that arise from expansion of magnetic 
field lines into the chromosphere. These jets are considerably 
larger and faster in the GOL simulation than in the non-GOL simulation. The jets 
in the GOL simulation shares many similarities with the so-called type II spicules. Their formation mechanisms and 
comparison with observations  
are investigated in more detail in a separate paper \citep{Martinez-Sykora:2017sci}. In addition to 
the similarities in their length ($\sim 8$~Mm) and speed ($\sim 100$~km~s$^{-1}$), the magnetic
field strength along the spicule also seems to be in accordance with observations
\citep[compare Figure~\ref{fig:1dstrat} with ][Figure
4]{Orozco-Suarez:2015kq}.

In short, the formation of many of these jets is caused by the expansion of magnetic
field into the chromosphere. When the ambipolar diffusion is large enough in the lower chromosphere and 
upper photosphere, the magnetic field can diffuse through the sub-adiabatic photosphere 
\citep{Leake2006}. We find that this occurs frequently in the vicinity
of strong flux concentrations when they interact with granular scale
fields in their vicinity. Once through the photosphere, the magnetic field will expand, producing a cold void characterized by low temperatures in the lower chromosphere 
(Figure~\ref{fig:histtgr})\footnote{ In fact, it is only
  in these regions in the GOL simulation that the ad-hoc heating term
  is playing some role. In contrast, the ad-hoc heating plays a key
  role in many more regions in the non-GOL simulation. To take this
  into account in the GOL simulation, we changed the threshold
  temperature for the ad-hoc heating term from 1600 to 1800 K at
  t=1460s (snapshot = 352). We note that the ad-hoc heating is below the
one needed for the non-GOL simulation (2000 K). This should be taken
into account while analyzing the simulation snapshots that are being provided.}.
In contrast, these weak field regions do not pass into the
chromosphere as often in the non-GOL simulation since it lacks
ambipolar diffusion.

\begin{figure*}
  \includegraphics[width=0.99\textwidth]{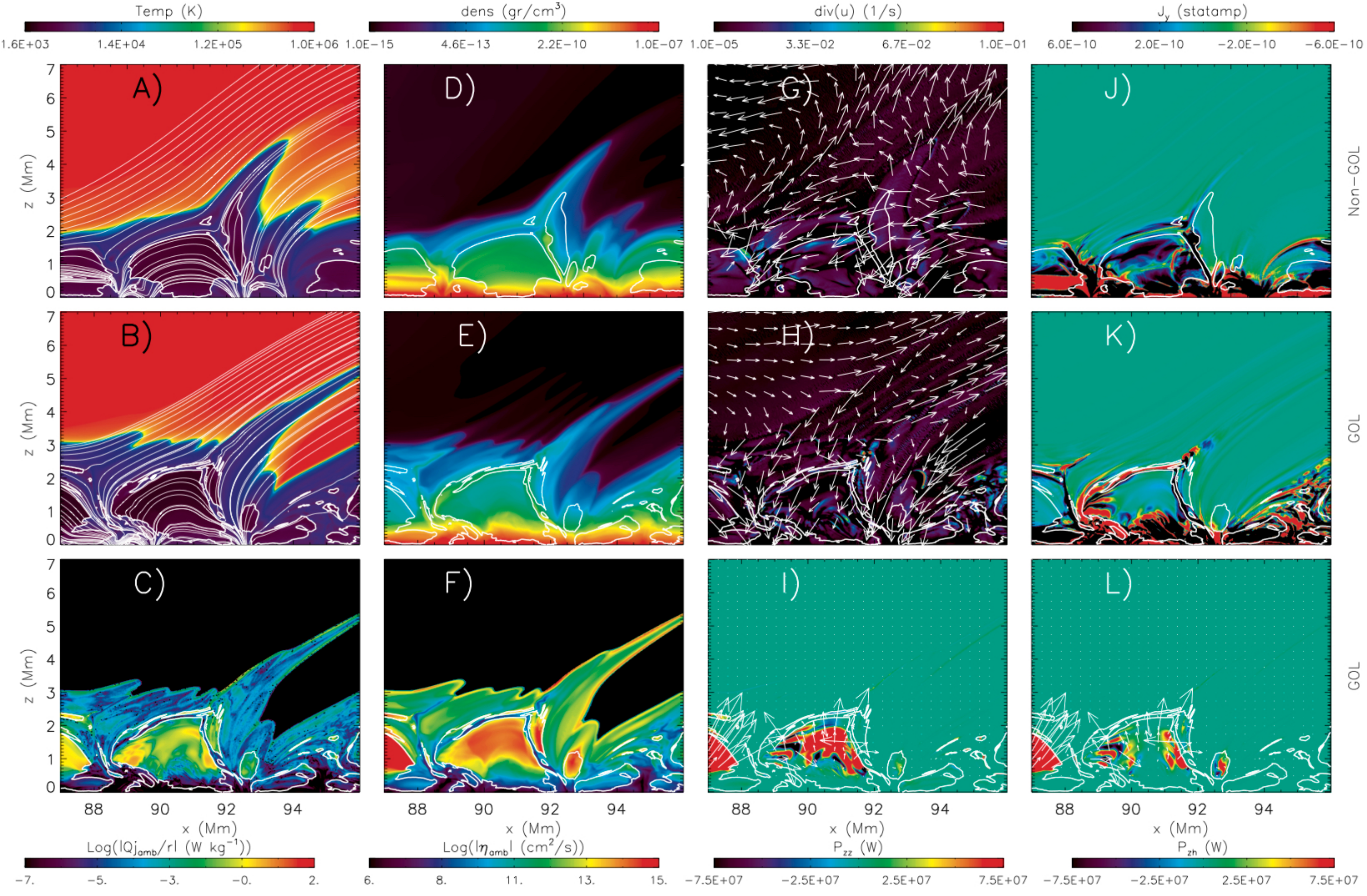}
  \caption{\label{fig:jetmaps} 
  Same layout as Figure~\ref{fig:shockmaps} for region  x=[87,96]~Mm except for panels D and E which 
 show the mass density at t=1708s (snapshot = 378) for the nonGOL and t=1360s (snapshot = 341) for the GOL simulation. }
\end{figure*}

The cold voids are a crucial step in the formation of the jets. As we describe
below, these voids are completely different from the cold bubbles that
occur in the wake of propagating shocks (described in section~\ref{sec:shocks}). 
Once again, ambipolar diffusion plays a critical role in the evolution of the cold
voids. The GOL simulation shows that the ambipolar diffusion is very high in these cold regions,
so that the diffusion occurs on short timescales ($\Delta t_A$). 
This means that the magnetic field in these cold voids 
diffuses on time-scales shorter than the lifetime of the voids.
In the non-GOL simulation this diffusion does not occur, 
and the nearly frozen-in magnetic field is significantly reduced as
the plasma expands. One example is shown in Figure~\ref{fig:jetmaps}: The expanding cold
void in the GOL simulation ($[x,z]\sim[91.5,1]$~Mm) shows fairly strong
currents and fairly strong
magnetic field. Magnetic field diffuses in the
cold void and the current is concentrated at the
exterior boundaries of the void (panel K in Figure~\ref{fig:jetmaps}). 

\begin{figure*}
  \includegraphics[width=0.99\textwidth]{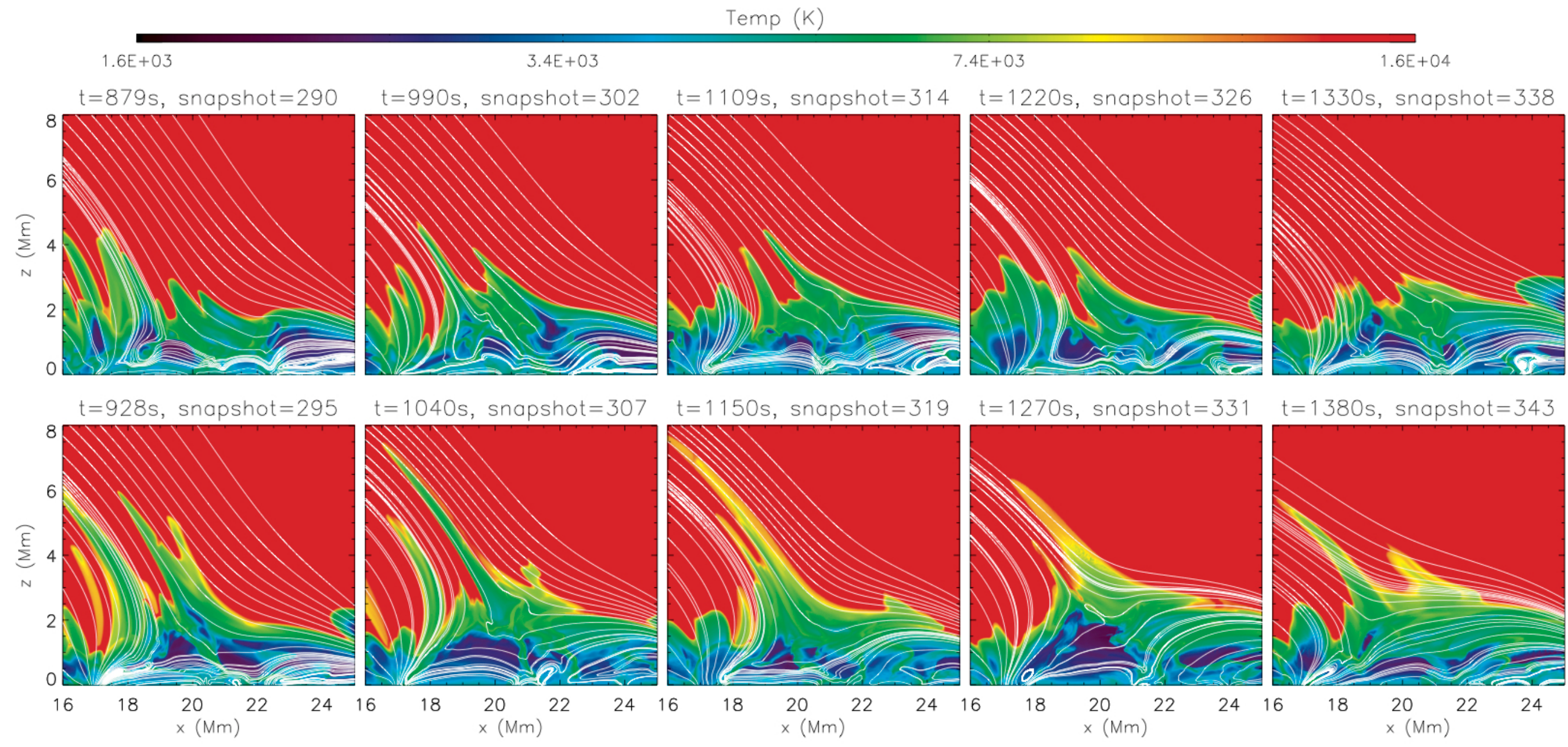}
  \caption{\label{fig:jetime} Time series of temperature maps showing spicule evolution in the 
 	non-GOL simulation (top) and the GOL simulation (bottom) in the region $x=[16,25]$~Mm. Magnetic field  
	is drawn with white lines (see the corresponding Movie 2). }
\end{figure*}

Upon expansion in the chromosphere, the magnetic field interacts with the ambient 
field which is inclined and penetrates into the corona.  It is the
release of the confined magnetic tension (confined because of the presence of 
ambipolar diffusion) that drives the strong flows that lead
to spicular jets.
As a result of these thin current layers, the magnetic
tension is much larger than in the non-GOL simulation. Therefore, 
more magnetic energy is released into kinetic and thermal energy 
(compare panels J and K in Figure~\ref{fig:jetmaps} and also the examples shown 
in Figure~\ref{fig:jetime}). 
Thus, with a larger magnetic tension, spicules  
outflows are longer and faster in the GOL simulation 
(see the jets that occur at the edges of the two plage polarities in
the Figure~\ref{fig:heat}).  

Figure~\ref{fig:jetime} shows temperature maps
for the non-GOL (top) and GOL simulations (bottom). 
There are three major differences between the simulations: 
1) the spicules are longer and faster in the GOL simulation; 2) magnetic field
lines seem to decouple from the thermodynamic structure of the spicule
near the end of their evolution in the GOL simulation (bottom right panel); 
3) and finally, the spicules are heated by ambipolar dissipation in the GOL
simulation and the temperature therefore increases with time. 
This temperature increase does not occur in the non-GOL simulation.

The heating of the spicules in the GOL simulation is of significant
interest since it matches well with observations of type II spicules
that seem to ``disappear'' in the
\ion{Ca}{2} passband but not in \ion{Mg}{2} or transition region lines 
\citep{Pereira:2014eu}. Detailed studies show that there is a clear
evolution of spicule strands being heated from chromospheric to
TR temperatures \citep{Skogsrud:2015qq}. The
temperature evolution of the spicules in the GOL simulation agrees
well with those observations \citep{Martinez-Sykora:2017sci}. 

The physical mechanism for heating the
spicules includes several components. Figure~\ref{fig:jetmaps} 
(from the GOL simulation) illustrates one component: the spicule shown carries current perpendicular to the plane of the simulation that is spread along the spicule axis. This current is
dissipated through ambipolar diffusion, heating the spicule to TR
temperatures despite the strong adiabatic expansion 
(see Movie~2 and the time-series in Figure~\ref{fig:jetime}). The rapid
expansion actually plays a key role in heating the spicular plasma: it
leads to a rapid drop in density so that the ambipolar
diffusion becomes larger (especially towards the spicule tops) since the ion-neutral
collision frequency is smaller. Since the spicule density is low compared 
to deeper regions in the chromosphere, ambipolar heating is so
effective that it can more than compensate for the adiabatic cooling
from the spicule expansion. There is also a second heating component:
there is additional heating through ambipolar
dissipation of the transverse waves that are driven by the release of the magnetic tension.
These two heating mechanisms explain the larger
density of points in the upper chromosphere and TR of
the 2D histogram of the GOL simulation in Figure~\ref{fig:histtgr}.  

The simulated type II spicules also impact the corona in
  a variety of ways. Shocks pass through, currents penetrate and heat
  the corona, flows fill the loops and thermal conduction spreads the resulting thermal energy through the million degree plasma in the associated loops. As a consequence of this, we see spicules impact various coronal observables. The impact of the type II spicules on the corona and the comparison with observations are described in detail in \citet{DePontieu:2017pcd} and Martinez-Sykora et al. (2017) (in prep).

\subsubsection{Chromospheric reconnection in highly inclined fields}~\label{sec:chromjets}

\begin{figure*}
  \includegraphics[width=0.99\textwidth]{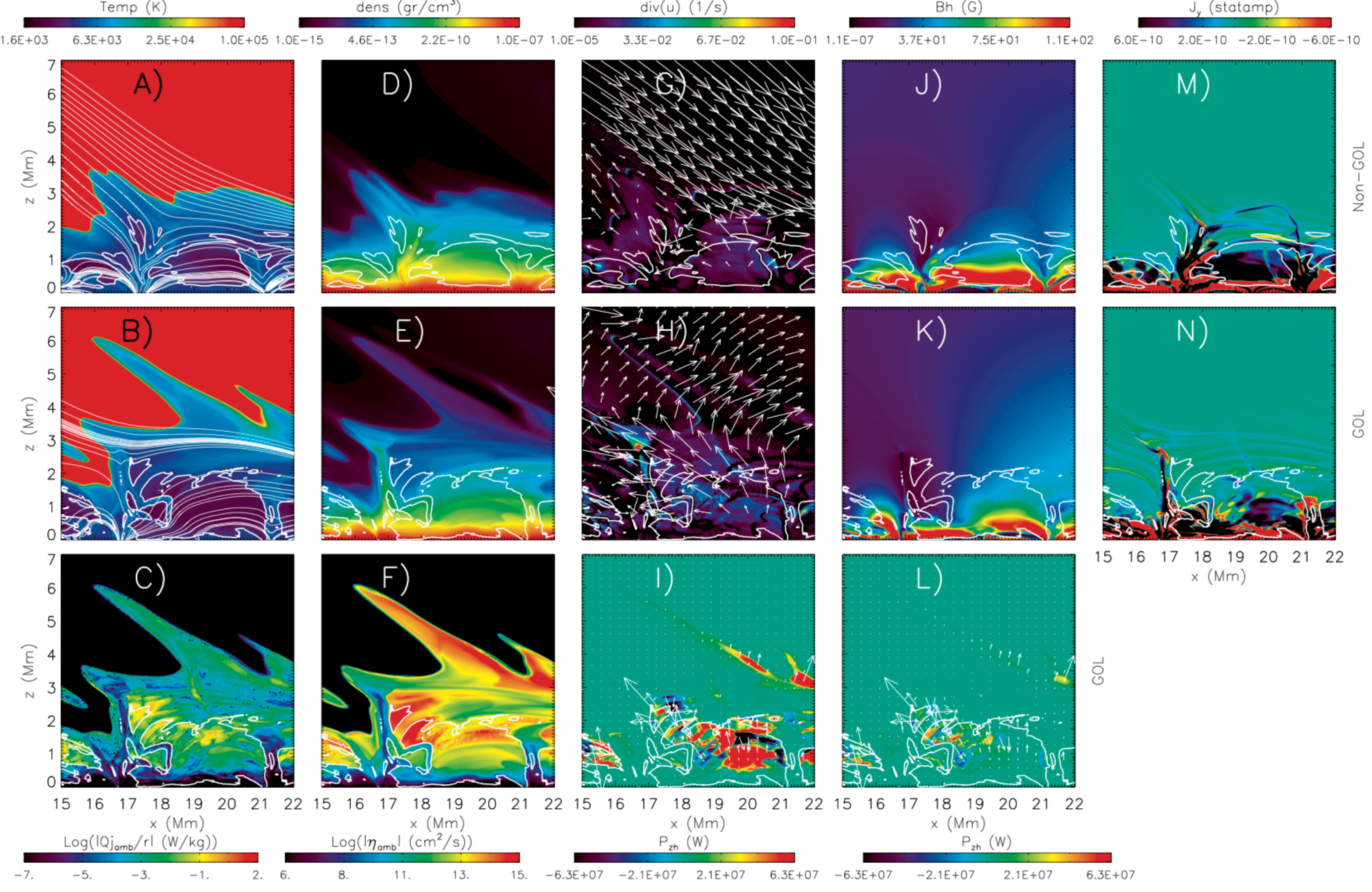}
  \caption{\label{fig:chromrecmaps} The top row shows, for the non-GOL
    simulation at t=1690s (snapshot = 376), maps of the temperature with white magnetic field lines (panel A), density (panel D),
    divergence of velocity, with velocity vectors in white (panel G),
    the horizontal magnetic field strength (panel J), and the electric
    current density perpendicular to the plane of the simulation
    (panel M). The middle row shows the same for the GOL simulation
     at t=1460s (snapshot = 352),
    with, respectively, panels B, E, H, K and N. The bottom row shows,
    for the GOL simulation,
    the Joule heating from the ambipolar diffusion (panel C), the
    ambipolar diffusion (panel F), and the vertical Poynting flux due to the vertical (panel I) and horizontal (panel L) ambipolar velocity. The white 
contours correspond to a $4000$~K temperature.
  The region shown is x=[50,57]~Mm and  x=[15,22]~Mm for 
  the non-GOL and GOL simulations respectively.}
\end{figure*}

Cold bubbles also form in the wake of shocks that propagate in regions
that include highly inclined (almost horizontal) magnetic field lines. In some
cases, this photospheric horizontal magnetic field may diffuse, {\it i.e.},
be advected with the help of the ambipolar velocity, into the middle-chromosphere 
(panels J, K, I, and L in Figure~\ref{fig:chromrecmaps}). This is similar to
what was described in Section~\ref{sec:jets}. 
Thus, the magnetic field expands into the outer atmosphere which
pushes chromospheric material to higher layers. With this magnetic field
topology, the magnetic field frequently forms dips in the proximity of 
intergranular lanes due to the downflows. 
As a result of these dips, the field often reconnects with ambient
field or magnetic field associated with neighboring cold bubbles (see panels A, B, M and N). 
Ambipolar diffusion enhances the reconnection rate as it concentrates
the current into thinner layers (panels F, I, L, M, and N). 
Since the magnetic field is highly 
inclined, this process does not create a jet into the corona but
rather helps to fill the upper chromosphere with plasma (panels D and E). 
As a result of these processes, magnetic energy is dissipated and transformed
into thermal energy in the upper chromosphere up to 
$5\,10^4$~K. We also see more kinetic energy in the GOL simulation in the 
upper-chromosphere, TR and lower corona as a result of the interaction of the diffused 
magnetic field with the ambient magnetic field. This leads to a wider range of densities in the 
upper chromosphere, TR and corona (Figure~\ref{fig:histtgr}). 

We note that the magnetic field topology where dips in the magnetic
field occur, shows some resemblance to the topology thought to play
a role in Ellerman bombs \citep{Georgoulis:2002qf}, except that in our case the magnetic field
strength is much weaker and the overlying fields are highly horizontal.

\section{Discussion and Conclusions}\label{sec:conclusions}

We have performed 2.5D radiative-MHD simulations of the solar atmosphere using 
the {\it Bifrost} code including ion-neutral interaction effects. These effects are implemented
in the code by adding the Hall term and the ambipolar diffusion (Pedersen dissipation)
in the induction equation including a new hyper-diffusive scheme to remove instabilities that may come from these two terms.

\citet{Martinez-Sykora:2012uq} described how the thermo-magnetic properties of 
the chromosphere lead to variations of the Hall term and 
ambipolar diffusion. In this paper we have detailed the impact of the partial ionization on the 
thermo-dynamics and magnetic field evolution. 

Before we summarize the differences between the simulations, we focus on 
a similarity between the two simulations which has not been reproduced 
in previous 2D self-consistent radiative MHD simulations. 
Despite the fact that the simulations are two dimensional, 
the corona is self-consistently maintained well above a million degrees. Previous 2D simulations 
required a hot plate at the top boundary in order to maintain a million degree corona 
\citep[{\it e.g.},][]{Leenaarts:2011qy,Heggland:2011kx,Iijima:2015fk,Nobrega-Siverio:2016qf}. 
It was necessary to expand into three dimensions and have closed field lines, {\it i.e.}, loops, 
in order to obtain a self-maintained hot corona 
\citep[{\it e.g.},][]{Gudiksen+Nordlund2002,Gudiksen:2005lr,Hansteen:2010uq,Martinez-Sykora:2011oq,Carlsson:2016rt}. 
Our two 2.5D simulations are different from previous 2D simulations:
they are higher resolution, include a larger numerical domain and
large-scale magnetic field connectivity. This combination of a large scale magnetic field and the large variety of physical 
processes and features that occur throughout the
simulated domain is key to produce a hot corona. Some of the key
processes that occur in our domain are: 1) magneto-acoustic shocks, {\it i.e.}, 
type I spicules with a large variety of magnetic field inclinations,
2) fast jets that appear similar to type II spicules, and 3) magnetic reconnection within the chromosphere where
the magnetic field lines are highly inclined. All of these processes combined 
build enough magnetic energy in the TR and low corona and are able to produce enough thermal energy to self-consistently maintain the corona at temperatures well above a million degrees.   
We also note that the large scale magnetic 
connectivity plays an important role in increasing the scale height of
the heating per particle compared to previous simulations: the heating
now reaches higher into the corona. This seems
to be in accordance with the results of \citet{Hansteen:2015qv}.
 
In addition to this similarity between the GOL (which includes partial
ionization effects) and non-GOL simulation,
there are many differences; from the chromosphere all the way 
into the corona. In the GOL simulation: 
\begin{itemize}
\item The expanding cold chromospheric bubbles are hotter due to the Joule heating
from the ambipolar diffusion when the bubbles reach low enough densities. 
\item The upper chromosphere and TR spread over a broader range 
of heights and densities. Similarly, the corona has a wider range 
of densities. 
\item The upper chromosphere is hotter since the ambipolar 
diffusion dissipates magnetic energy into thermal energy. This also
leads to spicules being heated during their evolution. 
\item The kinetic energy in the upper chromosphere, TR and 
lower corona is larger and shows a wider range of values 
as a result of more violent processes. 
\item The magnetic free energy in the upper-chromosphere and TR is larger than the non-GOL simulation by a factor of $\sim 1.3$ for the following reasons. 1) Some horizontal magnetic field that 
is normally constrained to the sub-adiabatic photosphere is diffused into the chromosphere. 
Consequently, the magnetic free energy in the photosphere is reduced
compared to the non-GOL simulation.  2) The increased
emergence into the chromosphere, combined with the interaction with
the ambient magnetic field leads 
to more magnetic free energy in the chromosphere. 
3) In addition, the boundaries between regions with high and low 
ambipolar diffusion often have strong variations of field line connectivity. 
\item In the corona, in contrast, the magnetic free energy 
is smaller since the simulated chromosphere 
is able to convert more magnetic energy into kinetic and thermal energy due to 
the ambipolar diffusion. 
\end{itemize}

Flux emergence plays a key role in the GOL simulation, even though we
did not impose explicit flux emergence at the boundaries of the
numerical domain. Ambipolar diffusion plays a key role in facilitating
the emergence of relatively weak magnetic flux that otherwise would
not be buoyant enough to expand into the atmosphere. We also found a
few locations where the Lorentz force points toward the convection
zone, so that the ambipolar diffusion, where it is large enough, helps to move the flux into
deeper layers. We found that once the flux has emerged into the chromosphere,
ambipolar diffusion often dissipates the currents introduced by the
emergence. We note that this does not always occur since it depends on
the magnetic field configuration as detailed in Section~\ref{sec:magnetic}. 

One of the most exciting aspects of the GOL simulation is that it is
the first Bifrost simulation in which jets that resemble type II
spicules are ubiquitously formed. These jets are formed by the release
of magnetic tension through a complex set of processes in which
ambipolar diffusion plays a key role. The resulting jets show very
high speeds, heights and thermal evolution that are similar to those
observed in type II spicules. The simulated jets occur mostly in the 
vicinity of the strong field regions, similar to what is observed on
the Sun. Ambipolar diffusion is required for the formation of these violent jets: it
allows fields to diffuse into the chromosphere, concentrates electrical currents into narrow regions that enhance the
magnetic tension, and it leads to the dissipation of currents and subsequent heating
of spicular plasma during the later phases of the evolution. Our
modeled spicules thus nicely reproduce the observational indications
for heating from chromospheric to TR temperatures during
the spicular lifetime \citep{Pereira:2014eu,Skogsrud:2015qq}. This
spicule formation process is further detailed in
\citet{Martinez-Sykora:2017sci}. The differences in spicule properties between the 
GOL and non-GOL model play an important role in explaining the different thermodynamic 
stratifications we have found in these models. 

Our simulations also show evidence of dynamics and mass loading that
is directly caused by reconnection (unlike the above-mentioned spicule
formation). For example, we find interesting dynamics in regions with
almost horizontal field lines above intergranular 
lanes where reconnection with neighboring cold bubbles leads to U shaped magnetic field lines. 
This type of reconnection does not produce jets that penetrate into
the corona, but it does help to fill the upper chromosphere with plasma. It is tempting to 
speculate that this process may play a role in providing mass to prominences. 
The role
of the ambipolar diffusion in this process is to 1) diffuse the magnetic
field narrowing the currents at the location of the reconnection, 2) lift 
more plasma into the upper chromosphere, 3) heat the cooler regions of the
chromosphere. 

Ambipolar diffusion also has another surprising effect: we find
that the magnetic field is sometimes decoupled from the thermo-dynamic 
structures. This occurs where ambipolar diffusion is large, 
the electrical current is perpendicular to the magnetic field lines, and the 
thermodynamic timescales are of the same order as 
the ambipolar timescales \citep[see][for
details]{Martinez-Sykora:2016qf}. This has (at least) three interesting
consequences. 

First, our GOL simulation shows that the field line connectivity often
changes during the evolution of spicules. This becomes particularly apparent towards
the end of the spicule lifetime.
This may provide an explanation to observations that suggest that
chromospheric fibrils do not necessarily follow the magnetic field
direction \citep{de-la-Cruz-Rodriguez:2011qd}.

Another consequence is
that it may invalidate modeling approaches that are based on tracking
the thermodynamic evolution along ``flux tubes'', i.e., 1D
hydrodynamic models. Clearly such models cannot properly capture the
thermodynamic evolution of plasma in regions where ambipolar diffusion
decouples the plasma from the magnetic field.

And finally, such decoupling may also affect magnetic field
extrapolation methods that incorporate the direction of chromospheric
features to constrain the extrapolation from photospheric
magnetograms.  Clearly such methods assume that
chromospheric features are aligned with the magnetic field
direction. Our simulation shows that this assumption breaks down
when the ambipolar diffusion is significant \citep[see][for
details]{Martinez-Sykora:2016qf}.
More generally, another challenge for such extrapolation methods is
the fact that the ambipolar diffusion in the chromosphere shows 
strong variations, so that regions with strong ambipolar diffusion may
be more potential, while at the boundaries between regions of 
strong and weak ambipolar diffusion the magnetic field lines may
undergo very significant changes in the magnetic connectivity. 

While these simulations show exciting results and potentially
significantly reduce the known discrepancies between Bifrost models and
observations of the chromosphere, several disclaimers are important to
note. While the presence of features that resemble type II spicules in
the GOL simulation may well reduce or even resolve long-standing
issues with chromospheric simulations, a proper comparison with
synthetic observables and an expansion of these results into three
dimensions is required to settle this issue. This is a major effort
that will be the subject of a follow-up paper.
Another issue that needs to be addressed in the future is the fact
that the current GOL simulation does not include time dependent
hydrogen or helium ionization, both of which will likely impact the
values and spatio-temporal distribution of the ambipolar diffusion.
 We would expect that the ambipolar diffusion 
gradients are reduced due to the time-dependent hydrogen and helium ionization 
\citep{Leenaarts:2007sf,Golding:2014fk}. In order 
to quantify the differences, future work needs to be based on
simulations that taking into account time-dependent hydrogen and helium ionization. 
Finally, the Generalized Ohm's Law is valid as long as 
the time scales are much larger than the ion-neutral collision
frequencies. It is not clear if this condition is always fulfilled
close to and in the TR. If this condition is not
fulfilled it may lead to drift between ions and neutrals \citep{Martinez-Sykora:2012uq}, which
requires a multi-fluid code to properly treat. This could alter some
of the results shown here. 

\section{Acknowledgments}

We gratefully acknowledge support by NASA grants NNX11AN98G,
NNM12AB40P, NNX16AG90G, NNH15ZDA001N-HSR, and NASA contracts NNM07AA01C (Hinode), and NNG09FA40C 
(IRIS). This research was supported  by the 
European Research Council under the European Union's Seventh Framework 
Programme (FP7/2007-2013) / ERC Grant agreement nr. 291058.
The simulations have been run on clusters from the Notur project, 
and the Pleiades cluster through the computing project s1061 from the High 
End Computing (HEC) division of NASA. We thankfully acknowledge the 
support of the Research Council of Norway 
through grant 230938/F50 and through grants of computing time from the 
Programme for Supercomputing. This work has benefited from discussions at 
the International Space Science Institute (ISSI) meetings on 
``Heating of the magnetized chromosphere'' where many aspects of this 
paper were discussed with other colleagues.
To analyze the data we have used IDL. We are very grateful to J. Vranjes for providing us the cross section tables as well as interesting conversations on subjects related to this paper. Finally, we want to thanks the referee for his/her help to improve the manuscript. 
\appendix
 
 \section{Data access}\label{sec:dataaccess}
 
 The full numerical domain with all the 
 variables \citep[listed in Table~2 in][]{Carlsson:2016rt} of the GOL simulation are freely available at the 
 {\it http://iris.lmsal.com/modeling.html} webpage, similar to
 the previously published enhanced network
 simulation \citep{Carlsson:2016rt}. For the current
 simulation, we also include the ambipolar diffusion, Joule heating coming from the ambipolar diffusion and electric current variables ($\eta_{amb}, q_{jamb}$, and $ix, iy, iz$, respectively).
 
 The current simulation has a different magnetic field
 configuration than the previously published enhanced network
 simulation. Since the magnetic field is the main free
 parameter of these realistic simulations, our current
 simulation thus complements the \citet{Carlsson:2016rt}
 simulation. In addition, the current simulation is only 2.5D
 but includes partial ionization effects. It does not include
 non equilibrium ionization, but the domain is larger and 
 with greater resolution. While using these models one has to
 take into account their limitations \citep[see below and
 ][]{Carlsson:2016rt}. In addition, one has to take into account
 that due to fast chromospheric expansion experienced in the
 GOL simulation (leading to cooler voids in the chromosphere), 
 we had to increase from 1600 to 1800 K the ad-hoc heating at t=1460s (snapshot = 352).
 
 Each snapshot is separated by roughly 10s and starts with t=0s at the snapshot number 200. The
 first available snapshot is at 280 which is 800s after the 
 partial ionization effects are switched on. The snapshots that
 are available cover the snapshots between 280 and 370, i.e.,
 $\sim13.8$ min. This time period includes the full 
 lifetime of several type II spicules including their impact on 
 the corona \citep[for details on this topic, see,and unpublished Martínez-Sykora et al. manuscript 2017]{DePontieu:2017pcd}. 
 
 Following the same format as for the Bifrost simulation that
 was previously made available \citep{Carlsson:2016rt}, the data are in FITS files with 2D cubes
 (x,z) with one variable for each file. The x-axis is equidistant and 
 can be generated using FITS keywords while the z-grid is non-uniform and is given in a FITS extension.
 
 The file names are of the form
 BIFROST\_en096014\_gol\_$<$var$>$\_$<$snap$>$.fits where the
 annotation en096014\_gol comes from ``enhanced network", 96~Mm
 and 14~km are the horizontal size, and grid-spacing, and
 ``gol" because the simulation includes ion-neutral interaction
 effects, i.e., the Generalized Ohm's Law. Similar to the
 previously published simulation, $<$var$>$ is the variable name listed in the search webpage, and $<$snap$>$ is the snapshot number.
 
 All variables have been cell centered on a right-handed system with z increasing upwards. Index runs the same way as the 
 axis which means that z[1] is at the bottom and z[nz] at the top. 
 
 All units are SI and given in FITS keywords (Mm, m~s$^{-1}$, kg m~s$^{-1}$, T, W~m$^{-3}$ , nm, T, etc.).  
 
 Metadata is given in the FITS header. This data release is
 part of the IRIS project and an explanation of the FITS
 keywords is given in IRIS Technical Note (ITN) 33. Software to
 analyze the simulation data is provided in SolarSoft
 (SSW/IRIS) with descriptions in ITN 34. Synthetic observables
 will also be made publicly available (see ITN 35). Papers
 published based on the simulation presented here should cite
 both the code description paper \cite{Gudiksen:2011qy} and the
 current paper.
 

\end{document}